\documentclass[12pt]{article}

\usepackage{lscape,multirow}
\usepackage{graphics,amsmath, amssymb,multirow, mathrsfs,graphicx,bm,colordvi,verbatim}
\usepackage{setspace, natbib}
\usepackage{caption, tabularx}
\usepackage{enumitem,colortbl}
\usepackage{url}
\usepackage{tablefootnote, ragged2e}
\usepackage{mathptmx}
\usepackage{pdflscape}
\usepackage{amsthm}
\usepackage{changepage}
\usepackage{tikz, tikz-cd, tikz-dependency}
\usepackage{float}
\usepackage[bottom]{footmisc}
\usepackage{eucal}
\usepackage{graphicx}
\usepackage{subcaption}
\usepackage{longtable}

\usepackage{hyperref}
\hypersetup{
    colorlinks=true,       
    linkcolor=blue,          
    citecolor=blue,        
    filecolor=magenta,      
    urlcolor=blue           
}

\usepackage{booktabs}

\newcommand{\sym}[1]{\rlap{#1}}
\newcommand{\citeay}[1]{\citeauthor{#1}, \citeyear{#1}}
\usepackage{dcolumn}
\newcolumntype{d}[1]{D..{#1}} 

\newcolumntype{Y}{>{\centering\arraybackslash}X}

\usepackage[left=1in,right=1in,top=1in, bottom=1in]{geometry}

\newcommand{\st}{\begin{eqnarray}}
    \newcommand{\nd}{\end{eqnarray}}
\newcommand{\stt}{\begin{eqnarray*}}
    \newcommand{\ndd}{\end{eqnarray*}}

\usepackage{setspace}
\usepackage{siunitx}

\graphicspath{{./figures/}}
\begin{document}


    \title{Do Sell-side Analyst Reports Have Investment Value?}
        \author{Linying Lv\thanks{Olin Business School, Washington University in St.\;Louis, llyu@wustl.edu.}}
        \date{\vspace{1cm}
        First draft: January 2025. This draft: August 2025.}   
        \maketitle
        \thispagestyle{empty}
        
    \begin{abstract}
    \begin{spacing}{1.2}
    \noindent This paper documents novel investment value in analyst report text. Using 1.2 million reports from 2000–2023, I embed narratives with large language models (LLMs) and fit machine learning (ML) forecasts of future long-term returns. Portfolios formed on the report narrative forecasts earn sizable and significant performance that is incremental to analysts' numerical outputs and to a broad set of established factors and characteristic‑based predictors. The effect is stronger after adverse news and is amplified for growth stocks with aggressive investment. To open the black box, I apply a Shapley decomposition that attributes portfolio performance to distinct topics. Analysts' strategic outlook contributes the most to portfolio performance, especially forward-looking fundamental assessments. Beyond providing direct evidence that analyst narratives contain value-relevant assessments that diffuse into price over time, this study illustrates how interpretable LLM-plus-ML pipelines can scale and augment human judgment in investment decisions.\\

    \end{spacing} 
    \begin{spacing}{1.2}
    \noindent \textbf{JEL Classification}: G11, G12, G14, G24\\
    \end{spacing} 
    \vspace{-0.1in}
    \begin{spacing}{1.2}
    \noindent \textbf{Keywords}: Analyst Report, Investment Value, Text Alpha, Large Language Models, Explainable AI\\
    \end{spacing} 
    \end{abstract}

\newpage
\setcounter{page}{1}

\onehalfspace

\section{Introduction}
\label{sec:intro}

In Q1 2020, Boeing experienced a significant stock crash, primarily due to the combined impact of the 737 MAX grounding and the COVID-19 pandemic. The stock plummeted as air travel declined and the company faced financial difficulties. From its February peak to mid-March, the stock fell by roughly 70\%. Sell-side analysts rapidly cut target prices and downgraded recommendations in response to the near-term collapse in demand.\footnote{For example, TD Cowen and Credit Suisse downgraded to Neutral on January 8, 2020, amid the MAX crisis. J.P.\ Morgan maintained an Overweight in January but downgraded to Neutral in March as the pandemic took hold.} Yet, reading the reports themselves reveals a more nuanced message: research houses emphasized constructive long-term fundamentals despite the near-term crisis. For example, J.P.\ Morgan regarded Boeing as ``among the companies best positioned to benefit from positive long-term aero fundamentals due to strong global demand'' and argued that ``Boeing's large backlog supports eventual recovery.'' Consistent with these narratives, Boeing's share price rebounded from about \$95 in mid-March 2020 to roughly \$240 by mid-March 2021.

This episode illustrates a critical gap in our understanding of how analyst reports create value. While existing literature has extensively documented the divergence in analysts' quantitative outputs (e.g., \citeay{malmendier2014security}; \citeay{birru2022analyst}), far less attention has been paid to the narratives that institutional investors consistently rank as the most valuable component of analyst research. These narratives serve a unique function: whereas numerical outputs typically focus on a one-year horizon, report narratives often provide crucial insights into longer-term fundamental assessments, risk factors, and the complex interplay between short-term headwinds and structural opportunities.

The Boeing case suggests that analyst narratives may contain investment-relevant information that is not fully captured in traditional quantitative measures. This raises important questions about how textual content in analyst reports contributes to market efficiency and whether narrative analysis can provide superior insights for investment decision-making. In this paper, I leverage the language ability of large language models (LLMs) to analyze 1.2 million analyst reports from Thomson Reuters Investext between 2000 and 2023 to test whether report narratives contain incremental investment value. I then apply a unique attribution framework to identify where the value originates.

I begin the analysis by examining the long-term return predictability of analyst report narratives. To capture the rich semantic content of these texts, I map each analyst report into a dense vector representation using LLM embeddings. These embeddings leverage the LLM's sophisticated linguistic understanding to convert unstructured textual information into high-dimensional numerical representations that preserve semantic meaning and contextual relationships.\footnote{This LLM-based approach to text representation has gained significant traction in recent finance studies (see, for example: \citeay{chen2022expected}; \citeay{chen2023deep}; \citeay{jha2025does}; \citeay{siano2025news}).} Using these text embeddings as input features, I employ machine learning models to extract value-relevant information and predict future 12-month stock returns. I then construct trading strategies based on these narrative-driven predictions to assess their economic significance.

The results demonstrate substantial predictive power across multiple return horizons. Analyst report narratives exhibit significant forecasting ability for future returns spanning 1 to 24 months. Stocks in the highest decile of narrative-based predictions generate average characteristic benchmark-adjusted returns of 4.7\% over the subsequent trading year, while those in the lowest decile experience average returns of -1.2\%, creating an economically meaningful spread of 5.9\% annually. This predictive power proves robust across several important dimensions. The results persist after controlling for revisions of recommendations, target prices, and earnings forecasts, suggesting incremental predictive power beyond what can be explained by analysts' numerical outputs. The predictive power also remains significant after controlling for traditional sentiment measures, indicating that LLM embeddings capture information beyond simple textual tone. Moreover, the findings are consistent across five different LLMs with distinct architectures and parameter specifications.

I next quantify the investment value of report narratives by constructing tradable portfolios. A narrative-based trading strategy generates substantial risk-adjusted returns. The portfolio delivers an average monthly return of 1.04\% with an alpha of 68 basis points (t-statistic = 2.64) relative to the \citet{fama2018choosing} six-factor model. The long-short strategy exhibits a low average monthly turnover of 28\%, indicating persistent signals that reduce implementation costs. The performance remains strong after accounting for transaction costs. To ensure these results are not driven by known return predictors, I conduct extensive robustness tests to benchmark against a strategy aggregating 94 characteristics-based factors from \citet{gu2020empirical} and 18 analyst-specific factors from \citet{ChenZimmermann2021}. Even after accounting for this comprehensive set of 112 established predictive factors, the narrative-based strategy maintains its strong performance, generating information ratios ranging from 0.73 to 1.41.

A potential concern in return prediction with pretrained LLMs is lookahead bias, where models inadvertently learn information about future market outcomes during pretraining \citep{sarkar2024lookahead}. I contend that my results are robust to this concern for three reasons. First, I embed the analyst report text using ChronoGPT$_{1999}$ and ChronoGPT$_{2024}$ from \cite{he2025chronologically}, models with identical architecture but knowledge cutoffs of December 1999 and December 2024, respectively. Under the lookahead bias hypothesis, ChronoGPT$_{2024}$ should substantially outperform ChronoGPT$_{1999}$ due to exposure to more recent market data. Instead, ChronoGPT$_{1999}$ achieves comparable performance (monthly return of 0.96\% with a Sharpe ratio of 0.71) to ChronoGPT$_{2024}$ (0.91\% return with a Sharpe ratio of 0.65), demonstrating that predictive power can emerge from genuine analytical capabilities without access to future market information. Second, I directly test for lookahead bias by examining portfolio performance after each model's knowledge cutoff date using four LLMs with training data ending between December 2018 and March 2023. If predictability resulted from information leakage, performance should deteriorate in post-cutoff periods. Instead, strategies generate superior performance with Sharpe ratios of 0.74 to 2.37, providing direct evidence that predictive power originates from genuine analytical capabilities. Third, if LLMs indeed learn and memorize future market information, we would expect to see strong predictability with the ``Company and Industry Overview'' content, as it provides the most comprehensive firm‑ and industry‑specific information. Instead, this content generates statistically insignificant alpha in attribution analysis.


Having shown that there is statistically significant and economically meaningful investment value in report narratives that is not driven by lookahead bias, I next investigate where and when the value originates. I first assess whether narrative-based predictions are especially informative for certain types of stocks. The evidence reveals substantial heterogeneity: return spreads between high and low decile predictions are significantly larger for growth stocks with aggressive investment policies. Consistent with this finding, factor regressions show the text-based strategy exhibits significant negative exposures to the value and investment factors, indicating stronger performance among growth-oriented firms with high asset growth. Furthermore, the strategy's performance is highly correlated with pro-cyclical industry portfolios such as business equipment and durables, suggesting that narrative-based predictions are particularly valuable for sectors sensitive to economic cycles and growth prospects. In contrast, predictive power shows minimal variation across analyst characteristics. The return spreads remain relatively stable across dimensions such as forecast accuracy, industry experience, coverage breadth, and brokerage firm size, suggesting that the narrative forecast's effectiveness is not concentrated among particular types of analysts but rather reflects collective wisdom generated by all analysts.

A potential concern is that narrative-based predictions simply capture the post-earnings-announcement drift (PEAD) effect documented by \citet{livnat2006comparing}, since analysts often issue reports around earnings announcements. To test for this channel, I examine the latest earnings surprise (SUE) prior to report release across narrative-based return forecast deciles. If my strategy were repackaging PEAD, stocks with high narrative-based return predictions should exhibit higher earnings surprises. The results reveal the opposite pattern. Stocks in the highest narrative return prediction decile are associated with the lowest earnings surprises, not the highest. This inverse relationship extends to broader market reactions: the two-day post-release abnormal returns are actually lower for stocks in the highest return prediction decile.

The evidence suggests a different mechanism that manifests in the Boeing case described above. Market overreaction to short-term negative news creates temporary mispricing despite solid long-term fundamentals, which is subsequently corrected. Supporting this interpretation, I find that analysts' numerical outputs are systematically more pessimistic for stocks with high narrative-based return predictions. Specifically, recommendations, target prices, and earnings forecasts are all lower in the highest narrative return prediction decile, consistent with analysts adjusting their quantitative assessments in response to near-term headwinds while their narratives capture longer-term value that the market temporarily underappreciates.


So far, I have shown that report narratives have investment value and that the value varies across stocks. Next, I turn my attention to identifying what content in analyst reports contributes to the value. I conduct an attribution analysis using the Shapley value decomposition framework following \citet{lv2024value}, designed specifically for LLM embeddings. To ensure economically meaningful results, I develop a comprehensive topic taxonomy. I organize report content into five mutually exclusive categories: (i) Company \& Industry Overview, (ii) Financial Analysis, (iii) Strategic Outlook, (iv) Risk \& Governance, and (v) Additional Content. I then generate return forecasts using different combinations of these topic categories and compute each category's Shapley value contribution to portfolio Sharpe ratios and returns.

The attribution analysis uncovers a striking concentration of investment value in one specific content type: Strategic Outlook. Although it makes up only 15\% of the average report, this section accounts for 41\% of the portfolio’s Sharpe ratio. In contrast, the far more prevalent Financial Analysis section contributes just 16\%, with other categories adding little or no value. Univariate portfolio performances verify the effectiveness of the attribution approach. A trading strategy using only textual embeddings of strategic outlook information earns 1.41\% per month (Sharpe = 0.93), while the others yield insignificant returns and alphas.

The economic content of the Strategic Outlook section centers on forward‐looking assessments of growth prospects, valuation, and investment opportunities that extend beyond short-term earnings projections. Consistent with the Boeing example, this content creates value by identifying long-term fundamentals temporarily overshadowed by short-term conditions. This pattern often occurs when quantitative downgrades are accompanied by qualitative statements of long‐term optimism. For example, in one representative case, an analyst wrote: ``Although the shares are an attractive long‐term investment, we must caution that the severity of the current industry downturn increases the chance of further near‐term earnings surprises.'' This language exemplifies the broader tendency observed across reports in which short‐term caution coexists with positive long‐term assessments. Further decomposition by timeframe, sentiment, and topic reveals that sentences emphasizing positive long-term fundamentals account for the majority of the investment value, confirming that analyst narratives create value by capturing forward-looking insights that temporary market pessimism obscures.

The collective evidence provides multiple lines of support for a mispricing interpretation over risk-based explanations. First, the narrative-based strategy generates substantial alphas that persist across four established risk factor models, suggesting the returns cannot be explained by compensation for systematic risk. Second, the concentration of investment value in forward-looking Strategic Outlook content, as revealed by the Shapley decomposition, points to information processing inefficiencies rather than risk characteristics that would be emphasized in Risk \& Governance categories. While I acknowledge that systematic risk may contribute, evidenced by factor loadings on growth and investment factors, the weight of evidence favors temporary mispricing that corrects as markets gradually incorporate the long-term insights embedded in analyst narratives.

This paper contributes to the extensive literature on sell-side analysts by demonstrating significant lookahead bias-free investment value in analyst report narratives that is incremental to established numeric outputs, including stock recommendations \citep{green2006value, christophe2010informed, chen2023revisiting}, earnings forecasts \citep{lys1990association, bordalo2019diagnostic}, and target prices \citep{brav2003empirical, farago2023analysts}. The study differs fundamentally from \citet{lv2024value}, who examines immediate price reactions around report releases. This paper studies long-horizon return predictability. The research questions and findings are therefore distinct: \citet{lv2024value} asks how quickly markets incorporate textual information and finds that Financial Analysis drives announcement-window reactions; this paper asks whether narrative content delivers persistent investment value and shows that Strategic Outlook language generates sustained abnormal returns.

Second, I add to the literature on textual analysis in asset pricing by implementing an LLM‐based topic‐level attribution framework. Distinct from prior work extracting sentiment or topics from corporate disclosures, earnings‐call transcripts, or news (e.g., \citeay{tetlock2007giving}; \citeay{manela2017news}; \citeay{chen2022expected}; \citeay{meursault2023pead}; \citeay{bybee2023narrative}; \citeay{bybee2024business}; \citeay{hirshleifer2023war}), my approach adapts the Shapley value decomposition method to systematically identify which specific themes in analyst narratives drive return predictability. I further validate the attribution by showing that a portfolio built solely on the highest‑contributing topic content delivers the strongest univariate performance.

Finally, I provide evidence consistent with a mechanism in which analysts create value by identifying long‐term fundamentals that are temporarily obscured by adverse short‐term conditions. This mechanism connects to \citet{bordalo2019diagnostic}, who examine long‐term growth expectations (LTG) and document behavioral biases in the overreaction to short‐term news. The distinguishing feature of the narrative signal, relative to LTG, lies in the opposite predictive relationship: whereas the LTG factor shows that long-term optimism predicts low returns (consistent with overreaction), my results demonstrate that narrative-based long-term optimism predicts high returns. This suggests that textual narratives provide more nuanced information relative to quantitative measures, capturing valuable forward-looking insights that quantitative analyst expectations miss. This finding also speaks to the long‐standing debate on whether analysts can select stocks for the long run (\citeay{altinkilicc2016can}; \citeay{birru2024analyst}). By exploiting the nuanced information embedded in narratives, I show that analyst research conveys two distinct signals: explicit recommendation changes and implicit judgments about long‐term prospects, with the latter carrying the investment value that persists over the long run.

The rest of the paper proceeds as follows. Section \ref{sec:data} describes the data and methodology, including the prediction models, topic classification approach, and Shapley value decomposition framework. Section \ref{sec:results} presents the empirical results. Section \ref{sec:conclu} concludes.

\section{Data and Methodology}\label{sec:data}
\subsection{Data}
I assemble the sample from two primary sources. From Mergent Investext, I collect 1,194,330 sell‑side analyst reports for S\&P 1500 constituent firms over 2000–2023. Table \ref{tab:sum1} reports annual counts for report characteristics, brokerage coverage, and unique analysts. I then obtain analysts’ numeric outputs—recommendations, EPS forecasts, target prices—and their announcement and revision histories from I/B/E/S.

To merge the datasets, I proceed in two steps. First, I map lead analysts in Investext to I/B/E/S analyst identifiers (AMASKCD) following \citet{li2024dissecting}. Second, I link each report to the corresponding I/B/E/S records within a five‑trading‑day window around the report date ([-2, +2]).

To benchmark against established predictors, I follow \citet{gu2020empirical} and construct 94 firm‑level characteristics for NYSE, AMEX, and NASDAQ firms. The sample spans January 1957 to December 2023, and monthly returns are from CRSP. Table \ref{anom} defines the characteristics.

To identify the incremental signal relative to quantitative analyst information, I compile 18 factors from \cite{ChenZimmermann2021} with significant investment values in analyst-produced information documented in the literature. These factors use analysts’ forecasts, recommendations, or coverage changes to derive signals that predict stock performance. Table \ref{ananum} provides a detailed description of the factors.

\subsection{Methodology}
\subsubsection{Prediction Models}

To capture narrative information, I quantify the perspectives embedded in analyst report text by mapping the report content into an embedding space. Compared with simple dimension tone measures in traditional NLP applications, the embeddings capture concise linguistic details and nuanced semantic meanings. Specifically, I input each report text into the LLaMA3-8B model and take the average of 32 transformer layer embeddings as text representation. 

To extract value‑relevant information, I use linear ridge regression to predict the next 12-month return following the report date. I deliberately use linear ridge regression models to emphasize the inherent information value of analyst reports rather than giving credit to sophisticated machine learning models. The ridge regression specifications are given by:
\vspace{-0.05in}
\begin{equation}\label{ridge_1}
Ret_{i t, 12m}=\beta_0+\beta^{\prime} y_{i j t}^{\mathrm{AI}}+\epsilon_{i j t},
\end{equation}
\begin{equation}\label{ridge_2}
\hat{\beta}=\underset{\beta}{\operatorname{argmin}}\left\{\left\|Ret_{i t, 12m}-\beta_0-y_{i j t}^{\mathrm{AI}} \beta\right\|_2^2+\theta\|\beta\|_2^2\right\},
\end{equation}
where $Ret_{i t, 12m}$ is the next 12-month return of stock $i$ following reports' release day $t$, and $y_{i j t}^{\mathrm{AI}}$ is the structured representation (4,096-dimension embeddings generated with the LLaMA3-8B model) of content from analyst report $j$ on stock $i$ released at day $t$. 

Specifically, I use a monthly expanding window. For each out-of-sample month $\tau$, I re-fit the ridge model on data available up to the prior month ($\tau-1$), and select the penalty via five-fold time-blocked cross-validation over a log-spaced grid. The fitted model is then applied to reports released in month $\tau$ to produce out-of-sample 12-month return forecasts. The initial training sample consists of the first five years of data (2000–2004), and the out-of-sample forecasting period runs from January 2005 to December 2023.

\subsubsection{Topic Modeling} \label{topic_modeling}

I implement a two‑step approach for interpretation. The first step assigns each sentence in the corpus to a semantically coherent and mutually exclusive topic. The second step attributes predictive power to these topics.

Standard unsupervised methods, such as Latent Dirichlet Allocation (LDA) or K-means clustering of embeddings, are ill-suited for this task due to two primary challenges. First, the repetitive and formulaic language of analyst reports causes these models to identify stylistic patterns rather than distinct economic topics. Second, unconstrained clustering tends to generate industry-specific topics based on sectoral jargon, creating clusters that are not comparable across the full sample of reports.

To address both issues, I impose a fixed taxonomy which is generated through systematically prompting following \citet{lv2024value}. The taxonomy specifies sixteen topics: \textit{Executive Summary}, \textit{Company Overview}, \textit{Industry Analysis}, \textit{Competitive Landscape}, \textit{Income Statement Analysis}, \textit{Balance Sheet Analysis}, \textit{Cash Flow Analysis}, \textit{Financial Ratios}, \textit{Business Segments}, \textit{Growth Strategies}, \textit{Risk Factors}, \textit{Management and Governance}, \textit{ESG Factors}, \textit{Valuation}, \textit{Investment Thesis}, and \textit{Appendices and Disclosures}. I add a residual category, \textit{None of the Above}, to capture boilerplate, data tables, and stray sentences that do not map cleanly to any predefined topic, yielding seventeen labels in total. 

I implement sentence-level classification with a fine-tuned BERT encoder, rather than a generative LLM, to exploit the stability and computational efficiency of supervised sequence classification at scale. The training set comprises 17,028 sentences from 100 randomly selected reports, labeled via an LLM-assisted protocol (ChatGPT-4o) with human verification. The trained model assigns each sentence a single topic label in $\{1, \ldots, 17\}$. On a stratified out‑of‑sample test set, the classifier achieves 89\% accuracy. Residual misclassification should behave like classical measurement error and therefore attenuate, rather than inflate, subsequent estimates. I then apply the classifier to the full corpus, labeling 52{,}350{,}385 sentences and constructing report‑level topic and meta‑category profiles for the downstream return‑prediction and attribution analyses.

I subsequently aggregate these fine‑grained labels into five meta‑categories that align with the main dimensions of sell‑side analyst research: company/industry overview, financial analysis, strategic outlook, risk and governance, and additional content. I aggregate the 17 fine‑grained topics into five meta‑categories for three reasons. First, interpretability: the meta‑categories map cleanly to the economic functions of sell‑side analyst research. Second, statistical power and robustness: aggregation reduces multiple testing and attenuates classification noise (89\% sentence‑level accuracy implies some residual error that averages out within broader groups). Third, econometric and computational tractability: attribution based on groups rather than many narrow topics yields stable Shapley estimates and avoids overfitting from correlated sub‑topics. 

Table \ref{tab: topic} describes each meta-category and its constituent topics. Figure \ref{fig: wordcloud} visualizes salient terms. The most prominent words in each category are consistent with the intended thematic distinctions. For instance, the Company and Industry Overview cloud emphasizes terms like ``growth'', ``market'', and ``products'', reflecting narrative content that introduces the firm’s position and operations. The Financial Analysis cloud highlights ``revenue'', ``sales'', ``estimate'', and ``EPS'', which are tightly linked to quantitative performance assessments. The Strategic Outlook cloud is dominated by action-oriented terms such as ``target'', ``stock'', ``believe'', and ``price'', mirroring analysts’ forward-looking investment views and valuation insights. In the Risk and Governance category, salient terms like ``risk'', ``management'', ``could'', and ``impact'' underscore discussions of uncertainty, decision-making, and potential downside scenarios. These distinct term profiles visually validate the semantic coherence of the meta-categories and reinforce their economic relevance for downstream analysis.


\subsubsection{Shapley Value Decomposition}

The second stage attributes the investment value of the report signal to the seventeen topics defined in Table \ref{tab: topic}. Because topics may interact, an attribution that ignores joint effects is potentially misleading: two topics can be weak in isolation yet informative in combination. I therefore use a Shapley value decomposition designed for LLM embeddings.

Specifically, I construct topic representations at the sentence level to limit attention across sentences. For report $j$ on firm $i$ at time $t$, let $y_{i j t d}$ denote the embedding of sentence $d_{\text {l}}$ and let $\mathcal{D}_{p, i j t}$ be the set of sentences assigned to topic $p$ by the classifier in Section \ref{topic_modeling}. I form the token-weighted topic vector
$$
\bar{y}_{p, i j t}^{\mathrm{emb}}=\frac{\sum_{d \in \mathcal{D}_{p, i j t}} \text {Tokens}_d  y_{i j t d}}{\sum_{d \in \mathcal{D}_{p, j i t}} \text {Tokens}_d}, \quad p \in\{1, \ldots, 17\},
$$
and then express the report-level embedding as a token-weighted average of topic blocks,
$$
y_{i j t}^{\mathrm{emb}}=\sum_{p=1}^{17} w_{p, i j t} \bar{y}_{p, i j t}^{\mathrm{emb}}, \quad w_{p, i j t}=\frac{\sum_{d \in \mathcal{D}_{p, i j t}} \text { Tokens}_d}{\sum_{q=1}^{17} \sum_{d \in \mathcal{D}_{q, i j t}} \text { Tokens}_d} .
$$

For any coalition $S \subseteq\{1, \ldots, 17\}$ of topics, I construct a representation that retains only the topics in $S$\footnote{I assign a zero block to the baseline feature value.},
$$
y_{i j t}^{\mathrm{emb}}(S)=\sum_{p \in S} w_{p, i j t} \bar{y}_{p, i j t}^{\mathrm{emb}}.
$$

Holding coefficients fixed, I generate coalition-specific predicted returns,
$$
\widehat{R E T}_{i j t}^{(S)}=\hat{\beta}_0+\hat{\beta}^{\top} y_{i j t}^{\mathrm{emb}}(S),
$$
and apply the same portfolio-formation rules as in the main tests (monthly sorts on $\widehat{R E T}_{i j t}^{(S)}$, value-weighted deciles, long-short $H-L$ constructed and evaluated over the target horizon). For each coalition, I compute a performance functional, $V_t(S)$, such as the next-period $H-L$ return or the annualized Sharpe ratio of the strategy constructed from $\widehat{R E T}^{(S)}$.


The Shapley contribution of topic $p$ in month $t$ is the average improvement in the performance functional as topic $p$ enters any subset $S$ that excludes it:
$$
\phi_{p, t}=\sum_{S \subseteq \mathcal{P} \backslash\{p\}} \frac{|S|!(P-|S|-1)!}{P!}\left[V_t(S \cup\{p\})-V_t(S)\right], \quad P=17, \quad \mathcal{P}=\{1, \ldots, 17\} .
$$

For interpretability, I aggregate 17 sub-topic level contributions to the five meta-categories introduced in Section \ref{topic_modeling}. Let $\mathcal{M}_g$ collect the topic indices belonging to meta-category $g \in\{1, \ldots, 5\}$. By additivity,
$$
\bar{\Phi}_g=\sum_{p \in \mathcal{M}_g} \bar{\phi}_p
$$
gives the contribution of meta-category $g$. This aggregation preserves exact additivity while yielding economically interpretable blocks.

\section{Empirical Results}
\label{sec:results}

\subsection{Quantifying Investment Value}

This section quantifies the investment value of analyst reports. Section \ref{pred} examines return predictability, while Section \ref{port} focuses on portfolio performance. Section \ref{alpha} assesses the incremental value beyond established asset pricing factors, and Section \ref{lab} addresses the potential lookahead bias concern.

\subsubsection{Return Predictability} \label{pred}

I begin by testing whether analyst narratives can predict future stock returns. I construct narrative-based predictions with the ridge regression \ref{ridge_1} and estimate the following report‑level regression:
\begin{equation}
RET_{i, t+12} = \alpha_{t} + \beta^{\prime} x_{i, t}+ \varepsilon_{i, t+12},
\end{equation}
where $RET_{i, t+12}$ is the realized 12‑month return of stock $i$. $x_{i, t}$ denotes distinct signals from each report, consisting of four measures: $\widehat{RET}_{12m}$ is the text‑based 12‑month return forecast; $REC_{REV}$ represents recommendation revision, calculated as the difference between the current report's recommendation and the analyst's last recommendation for the same stock in I/B/E/S; $EF_{REV}$ denotes earnings forecast revision; and $TP_{REV}$ represents target price revision. $Tone$ assesses report sentiment, using a fine-tuned BERT model that classifies each sentence as positive, negative, or neutral. Specifically, $Tone$ is calculated as:
\begin{equation}
Tone_{i}=\frac{N_{i}^{+}-N_{i}^{-}}{N_{i}},
\end{equation}
where $N_{i}^{+}$ ($N_{i}^{-}$) represents the number of positive (negative) sentences in report $i$ as classified by the BERT classifier, and $N_{i}$ denotes the total number of sentences in the report. The specification includes year-month fixed effects to absorb common time-series variation, ensuring that the coefficients capture cross-sectional predictive power. Standard errors are clustered by both firm and year-month.

Table \ref{tab:predictability} reports the regression estimates. Panel A shows a pronounced, nearly linear relation between the text-based forecast $\widehat{R E T}_{12 m}$ and realized returns from one to twenty-four months. Each unit increase in $\widehat{R E T}_{12 m}$ raises the next-month return by 0.30 percent ($t=1.85$). The effect grows to 0.60 percent for the twelve-month window ($t=4.37$) and 1.14 percent for the twenty-four-month horizon ($t=5.24$). This monotonic pattern suggests a gradual, rather than instantaneous, incorporation of the narrative's information into market prices.

Panel B contrasts the text-derived signal with standard numerical revisions. In univariate specifications (columns 1-4), none of the numerical outputs, like recommendation revisions, earnings forecast revisions, or target price revisions, nor a simple sentiment score, exhibits significant predictive power. By contrast, the narrative forecast $\widehat{R E T}_{12 m}$ is a strong predictor on its own (Column 5). Crucially, it retains its economic and statistical significance when all other signals are included as controls (Columns 6-7). In the fully specified model with year-month fixed effects, the coefficient on the narrative signal is 0.10 (t=3.34), and it remains highly significant even with the inclusion of more stringent fixed effects. In all multivariate specifications, the predictive power of quantitative revisions and the $Tone$ score is statistically and economically negligible.

Taken together, these findings indicate that analyst narratives contain valuable, long-horizon information that is priced slowly. Furthermore, the predictive power of the narrative-based signal is distinct from, and dominates, that of analysts' quantitative outputs.

To visualize the return dynamics following report issuance, I form portfolios based on daily sorts. Each day, all reports released are sorted into deciles based on their out-of-sample return predictions. I then track the subsequent performance of these portfolios by calculating their buy-and-hold abnormal returns (BHAR) using the characteristics-based benchmark approach of \citet{daniel1997measuring} (DGTW). The cumulative abnormal return (CAR) over a T-day horizon is defined as:
\begin{equation}
C A R_{0, T}=\prod_{t=0}^T\left(1+R_{i t}\right)-\prod_{t=0}^T\left(1+R_{i t}^{D G T W}\right),
\end{equation}
where $R_{i t}$ represents the raw return of stock $i$ on trading day $t$, $R_{i t}^{D G T W}$ denotes the value-weighted return of stock $i$'s benchmark portfolio on day $t$, and $T$ indicates the end date of the accumulation period. $t=0$ represents the first trading day at or after a report is released.

\begin{figure}[!htb]
\captionsetup{skip=-0.5em}
\caption{Report-based Forecasts and Abnormal Stock Returns} \label{fig:car}
\begin{center}
	\includegraphics[width = \textwidth]{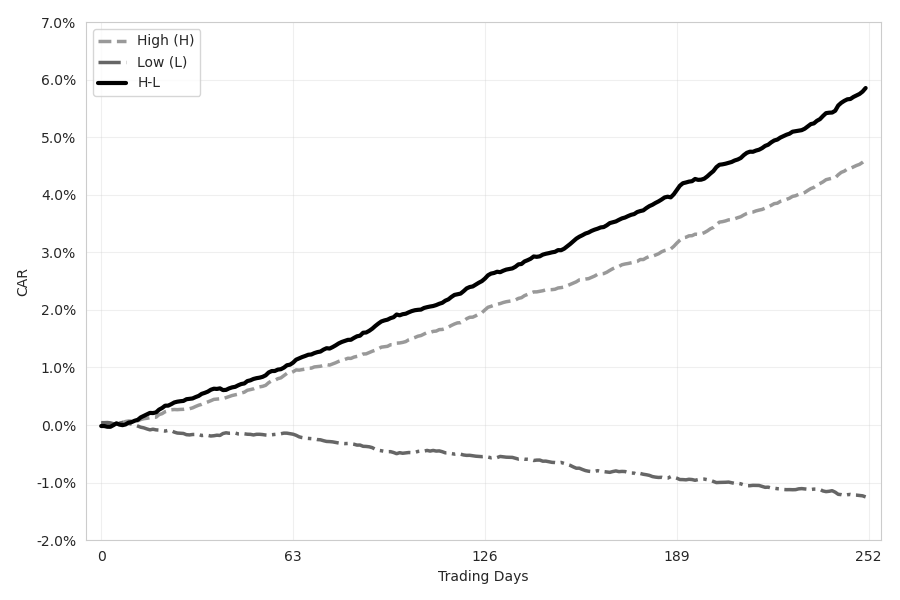}
\end{center}
\vspace{-0.2in}
{\footnotesize Note: This figure presents DGTW characteristic-adjusted buy-and-hold abnormal returns following research report releases, measured over a [0, +252] trading day window after each announcement. Reports are sorted into deciles based on their predicted 12-month returns. The `High (H)' group consists of reports predicting returns in the highest decile within each trading day, while the `Low (L)' group comprises reports predicting returns in the lowest decile. Returns are value-weighted within each calendar month using lagged market-capitalization weights and then equally averaged across months. `H-L' denotes the return spread between the High and Low groups. The sample spans from 2005 to 2023.}
\end{figure}

Figure \ref{fig:car} plots the average cumulative abnormal returns for the top decile (most positive predictions) and bottom decile (most negative predictions) portfolios over the subsequent 252 trading days. The figure shows a clear and persistent divergence between the two portfolios. The top-decile portfolio earns a cumulative abnormal return of 4.7\% over the year, while the bottom-decile portfolio underperforms its benchmark by -1.2\%. The resulting long-short spread of 5.9\% widens steadily over the period, demonstrating that the market gradually incorporates the long-horizon information contained in the report narratives.

To ensure the robustness of these findings, I verify the results using four diverse language models spanning different scales, architectures, and training periods: BERT (110M parameters), RoBERTa (355M parameters), LLaMA2-13B, and ChronoGPT$_{1999}$, a chronologically consistent model developed by \citet{he2025chronologically} that was trained exclusively on text timestamped prior to December 1999, ensuring complete isolation from my sample period. Figure \ref{fig:car_alterllm} confirms that the predictive pattern persists across alternative language models.

Table \ref{tab:ls_stat} presents the statistical significance tests for the ``High'', ``Low'', and ``High-Low'' portfolios. For each group, I calculate the value-weighted average of the cumulative abnormal returns over 252 trading days and test for significance using Newey–West standard errors adjusted with 12 lags to account for serial correlation. The results show that all models generate statistically significant long-short spreads. LLaMA3 achieves the highest spread of 5.9\% (t = 3.42), followed by LLaMA2 at 5.0\% (t = 3.10) and ChronoGPT$_{1999}$ at 3.1\% (t = 2.82). Even the smaller models, BERT and RoBERTa, produce significant spreads of 2.1\% and 1.9\%, respectively. These results confirm that the predictive power of narrative sentiment is robust across different model architectures and is not an artifact of any particular language model's training or design.

\subsubsection{Performance of Long-Short Portfolios} \label{port}

I proceed with a portfolio analysis to evaluate the economic value of analyst report predictability. The trading strategy design incorporates two key elements. First, portfolios are rebalanced monthly to balance new information and transaction costs. Second, motivated by the persistence in return predictability documented in Panel A of Table \ref{tab:predictability}, I aggregate analyst report forecasts over extended historical windows to capture sustained signals in analyst reports.

Specifically, at the beginning of each month, I construct portfolios with the following procedure. First, for each stock, I calculate the average predicted return from all analyst reports issued during the past $L B$ months, where $L B \in\{9,12,18,24\}$. I then sort stocks into decile portfolios based on these averaged predictions. Within each decile portfolio, stocks are weighted by their previous month-end market capitalization to mitigate the influence of microcaps. I then construct a zero-cost long-short portfolio that takes a long position in the highest predicted return decile (decile 10) and a short position in the lowest predicted return decile (decile 1).

Table \ref{tab:decile} presents performance statistics of value-weighted decile portfolios with different lookback periods. Across all $LB$, the decile spread is economically large and statistically significant: the $H-L$ portfolio earns $0.87 \%-1.16 \%$ per month with Sharpe ratios of $0.62-0.69$. Risk adjustment using the Fama-French six factors yields positive, significant alphas of $0.52 \%-0.79 \%$ per month (t-statistics between 1.99 and 2.64). The cross-decile pattern of alphas indicates that performance is primarily driven by the long leg rather than the short leg, consistent with the asymmetric price paths in Figure \ref{fig:car}. Based on this evidence, I use the 12-month lookback window as the benchmark in subsequent analyses: it delivers the highest $H-L$ alpha t-statistic (2.64) and aligns with the model's 12-month return-prediction horizon.

The extended lookback windows effectively smooth analyst signals and substantially reduce portfolio turnover, demonstrating the persistent nature of analyst predictions. Table \ref{tab:txcost_robust} reports turnover statistics for the long leg, short leg, and combined long-short portfolio across different lookback periods.\footnote{Appendix B details the implementation of portfolio turnover and transaction costs.} The results reveal a clear inverse relationship between lookback window length and turnover rates. Specifically, long-short portfolio turnover decreases from 34.0\% for the 9-month lookback to 18.4\% for the 24-month window, a reduction of nearly half. This declining pattern holds consistently across both individual legs, with long-leg turnover falling from 26.4\% to 13.7\%, and short-leg turnover decreasing from 41.6\% to 23.2\% over the same range.

The substantial reduction in turnover rates with longer lookback windows provides compelling evidence for the persistence of analyst signals embedded in research reports. Lower turnover not only confirms signal stability but also creates significant economic benefits through reduced transaction costs. When accounting for realistic transaction costs, the trading strategy remains highly profitable across all specifications. Even under the more conservative 60 basis points cost assumption, representing higher trading costs for small-cap stocks as documented in \citet{demiguel2020transaction}, the strategy generates statistically significant net monthly returns ranging from 0.579\% (9-month lookback) to 0.947\% (24-month lookback). Under the 35 basis points cost scenario, reflecting large-cap trading expenses, net returns are even more attractive, ranging from 0.751\% to 1.040\% monthly.

\begin{figure}[!htb]
\captionsetup{skip=-0.5em}
\caption{Cumulative Log-return of Portfolios Sorted on Report Forecasts}
\begin{center}
\label{fig:cum_ret}
	\includegraphics[width = 0.9\textwidth]{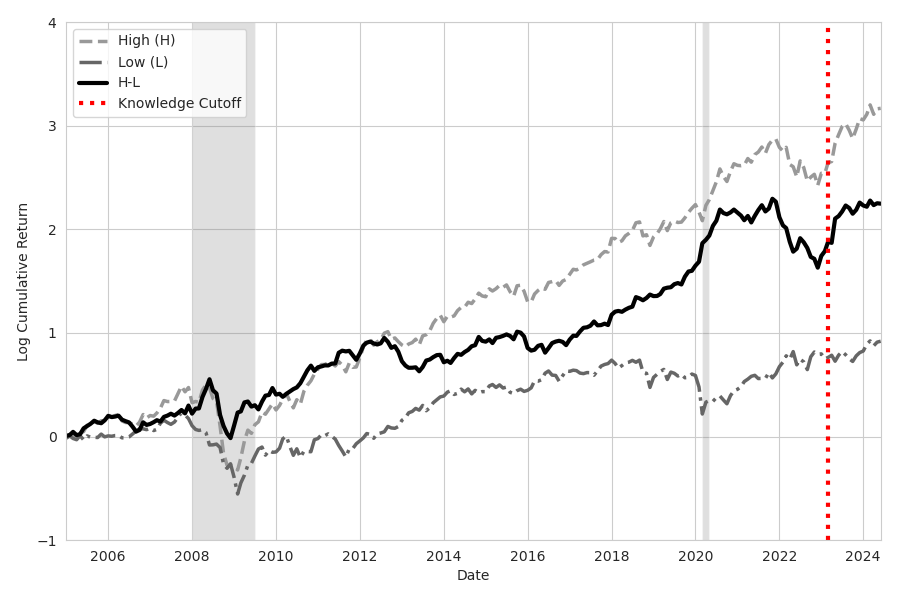}
\end{center}
\vspace{-0.2in}
{\footnotesize Note: This figure shows cumulative log excess returns for portfolios constructed based on analyst report forecasts from the previous 12 months. For each panel, three portfolios are presented: the highest decile (H), the lowest decile (L), and a long-short portfolio (H-L) that buys the highest and sells the lowest decile. Grey shaded areas represent NBER recession periods. The dashed vertical line marks the LLaMA3-8B model knowledge cutoff. The return time series spans from January 2005 to June 2024.}
\end{figure}

Figure \ref{fig:cum_ret} plots cumulative returns for the benchmark $LB=12$ portfolio from January 2005 to June 2024. The self‑financing long–short portfolio compounds to 846\% over the sample, compared with a 399\% cumulative excess return on the value‑weighted market. Decomposing the legs, the long side compounds to 2,274\% over the period, while the short side delivers 151\%. Thus, the investment value stems from picking stocks with strong future prospects, which is consistent with the portfolio statistics in Table \ref{tab:decile}. The strategy experiences significant drawdowns during the global financial crisis and again post-2022, conditions in which fundamental signals can be temporarily overshadowed by systemic risk aversion and forced liquidations.

\subsubsection{Incremental Investment Value} \label{alpha}

The next question is whether the predictive signals from analyst reports provide incremental value or merely reflect exposure to known risk factors. I address this by computing risk-adjusted returns ($\alpha$) against four benchmark models: the \citet{fama2015five} five factor model ($\alpha_{F 5}$), the \citet{fama2018choosing} six factor model ($\alpha_{F 6}$), the q -factor model of \citet{hou2015digesting} ($\alpha_{H X Z}$), and the behavioral factor model of \citet{daniel2020short} ($\alpha_{DHS}$). I then conduct a broader analysis to see if the report signals contain information beyond the 94 characteristics-based factors documented in \citet{gu2020empirical}.

I also examine whether the narrative-based signals are informative relative to traditional quantitative signals produced by analysts. To do this, I test for incremental value over 18 established analyst-based factors, such as EPS forecast revisions and recommendation changes, from the \citet{ChenZimmermann2021} database.

In Table \ref{tab:IIV}, I evaluate the predictive power of analyst reports relative to and in combination with established return predictors. The analysis examines three main signal families: analyst report narrative predictions (RP), the $1/N$ average return of 18 analyst-based factors (ANA), and predictions from a 4-layer artificial neural network (ANOM), the best-performing machine learning model in \citet{gu2020empirical}. For each strategy and combination, I report mean returns, Sharpe ratio (SR), and information ratio (IR). The Information Ratio of the portfolio $i$ relative to a benchmark $b$ is calculated as:
\begin{equation}
IR_{i,b} = \frac{\bar{r}_{i,b}}{\sigma_{i,b}},
\end{equation}
where $\bar{r}_{i,b}$ is the average excess return (difference between portfolio $i$ and benchmark $b$ returns) and $\sigma_{i,b}$ is the standard deviation of these excess returns. In Table \ref{tab:IIV}, I compute the IR of combined strategies relative to individual components to measure the incremental value of information sources. For example, the IR of ``RP + ANA" measures the incremental gain from combining report predictions with analyst-based factors, relative to using ``ANA'' signals only.

Panel A first establishes the standalone performance of each information source. Report text generates significant $\alpha$ across various factor models, with alphas ranging from 0.73 to 1.21. 
The substantial and consistent alpha across multiple established risk factor models suggests the narrative-based strategy captures mispricing rather than compensation for systematic risk. If the returns reflected risk premiums, we would expect the alphas to diminish substantially when controlling for comprehensive risk factors. Instead, the persistence of economically and statistically significant alphas across the Fama-French five- and six-factor models, the q-factor model, and behavioral factors indicates that traditional risk-based explanations cannot fully account for the results.

The analyst-based predictors (ANA) show modest but significant predictive power (alphas between 0.16 and 0.28). The fundamental-based anomaly portfolio (ANOM) exhibits strong performance with alphas between 1.15 and 1.40.

Panel B addresses the central question of incremental value. When combining report predictions with analyst-based factors (RP + ANA), the portfolio maintains significant alphas (0.50-0.68) and shows meaningful incremental gain with an IR of 0.73. Most notably, incorporating both analyst and fundamental factors (RP + ANA + ANOM) produces the highest Sharpe ratio (1.60) and consistent alphas across factor models (0.77-0.92), with an IR of 1.23 (t=4.05).

These results suggest that analyst reports contain statistically and economically meaningful incremental information not captured by a broad range of analyst-based and fundamental-based predictors. The significant incremental performance and information ratios indicate that textual information in research reports offers a distinct and complementary source of investment value.

\subsubsection{Lookahead Bias} \label{lab}

A critical concern in demonstrating out-of-sample profitability is the potential for lookahead bias in LLMs. Since these models are pretrained on massive text corpora, they may have indirectly absorbed information about future stock performance, creating spurious predictive ability.

To address this concern directly, I conduct a stringent analysis using only post-knowledge-cutoff samples of the following models: BERT (110M parameters), RoBERTa (355M parameters), LLaMA2-13B, and LLaMA3-8B. The testing methodology is straightforward yet stringent: for each model, I construct portfolios and evaluate performance exclusively during periods following the model's knowledge cutoff date. This approach creates genuine out-of-sample conditions by ensuring complete temporal separation between training data and evaluation periods.

Second, I employ ChronoGPT models from \citet{he2025chronologically} to directly test for lookahead bias across the full sample period.\footnote{\citet{he2025chronologically} release two model families: ChronoBERT and ChronoGPT. I select ChronoGPT due to its larger context length (1,792 versus 1,024 tokens), which better accommodates analyst reports averaging approximately 1,300 tokens.} These models share identical architecture and parameters but differ in knowledge cutoffs: $\text{ChronoGPT}_{1999}$ (cutoff: December 1999) and $\text{ChronoGPT}_{2024}$ (cutoff: December 2024). Under the lookahead bias hypothesis, $\text{ChronoGPT}_{2024}$ should substantially outperform $\text{ChronoGPT}_{1999}$ due to exposure to more recent market information.

Table \ref{tab:lookahead} reports the results of this temporal validation. Remarkably, all models generate substantial positive H-L returns, ranging from 0.91\% ($\text{ChronoGPT}_{2024}$) to 2.70\% (LLaMA3), with corresponding Sharpe ratios between 0.65 and 1.28. Crucially, $\text{ChronoGPT}_{1999}$ achieves a comparable ``H-L'' return (0.96\%) and Sharpe ratio (0.71) compared to $\text{ChronoGPT}_{1999}$ (0.91\% and 0.65, respectively). This pattern directly contradicts the lookahead bias hypothesis, which would predict superior performance from the model with the more recent knowledge cutoff.

These findings provide compelling evidence that the documented profitability stems from genuine information extraction from contemporaneous analyst narratives, rather than memorized future outcomes embedded in pretraining corpora.

\subsection{Where Does the Investment Value Come From?} 

I next ask what types of analysts and firms are associated with high return forecasts. Section \ref{xs} characterizes the cross‑section of signals. Section \ref{news} studies state dependence conditioning on contemporaneous news.

\subsubsection{Which Stock is Picked?} \label{xs}

Table \ref{tab:chars} presents a cross-sectional analysis, comparing the characteristics of firms and analysts in the top and bottom deciles of the narrative-based return forecast $\widehat{RET}_{12m}$. The table reports means and medians for each group, along with t-tests and Wilcoxon rank-sum tests for differences between the high and low forecast groups.

Panel A indicates limited dispersion in analyst attributes. While reports with high forecasts are issued by analysts with slightly narrower coverage (fewer industries and firms), and from smaller brokerages, the economic magnitudes of these differences are small. Other key attributes, such as analyst experience and affiliation with a ``Top 10'' brokerage, show no meaningful difference across the groups. This pattern suggests the narrative signal is not simply a proxy for analyst status or coverage breadth.

Panel B reveals that the signal has a mild tilt toward certain firm characteristics. Firms with high text-based forecasts tend to be larger, more profitable growth stocks. Compared to firms in the low-forecast group, these ``High'' firms have a higher market capitalization (mean log market cap 16.18 vs. 16.06), lower book-to-market ratios (0.37 vs. 0.45), and higher gross profitability (0.39 vs. 0.35). Investment is higher among ``High'' return prediction groups (1.01 vs. 0.98), while idiosyncratic volatility is modestly higher for the High group (0.02 vs. 0.01). Again, while these differences are statistically significant by both t‑tests and Wilcoxon rank‑sum tests (see p‑values in the table), their magnitudes are economically small.

The correlations between return forecasts and characteristics could be incidental, or they could indicate that the signal's predictive power is concentrated in certain types of firms or analysts. To distinguish between these possibilities, I analyze how the forecast's performance varies across subsamples sorted on firm and analyst attributes. Specifically, the sample is first partitioned by the monthly median of a firm-level or analyst-level characteristic (``Above'' vs. ``Below''), and then within each group, I trace the average abnormal returns for portfolios formed on the top and bottom deciles (``High'' vs. ``Low'') of the narrative-based return forecasts over a 252 trading day horizon. I then compare the difference in spreads across ``Above-High'' minus ``Above-Low'' and ``Below-High'' minus ``Below-Low''. If there is a significant difference in the spread between subsamples, it suggests that the predictive power varies systematically with the underlying characteristic.

Table \ref{tab:char_stats} presents the statistics for this conditional analysis. Panel A reveals that analyst characteristics show limited influence on the signal's predictive power. The long-short spreads are relatively similar across analyst attributes such as forecast accuracy, industry coverage breadth, firm coverage, experience, top-tier brokerage affiliation, and broker size. The differences in spreads between above-median and below-median analyst groups are generally small and often statistically insignificant, suggesting that the narrative signal's effectiveness is not primarily driven by analyst quality or institutional affiliation.

Panel B shows more pronounced variation across firm characteristics. The signal demonstrates stronger predictive power for certain types of companies. Large firms (above-median market capitalization) generate significantly higher long-leg returns (5.0\% vs. 2.6\% for small firms), with a statistically significant difference in spreads of 3.0\%. Growth stocks (below-median book-to-market) also exhibit superior predictive performance compared to value stocks. High-profitability firms show stronger signal effectiveness, with the spread difference reaching statistical significance. The investment characteristic also reveals meaningful variation, with high-investment firms displaying greater sensitivity to the narrative signal. Finally, idiosyncratic volatility shows that high-volatility stocks generate larger spreads, though the difference is insignificant.

These traits are mirrored in a regression on the \citet{fama2018choosing} six factors. Table \ref{tab:alpha_ff} shows that the ``H-L'' portfolio is close to market-neutral but not flat (market beta $\approx$ 0.27). Size exposure is near zero. Loadings on value and investment are strongly negative, consistent with a growth and aggressive-investment tilt. The profitability loading is positive but modest, and momentum is statistically weak. The long-only leg exhibits the corresponding pattern: a high market beta ($\approx 1.20$), negligible size exposure, pronounced negative HML and CMA loadings. Taken together, the cross‑sectional comparisons and factor regressions imply that the text‑based forecasting power primarily originates in growth firms with aggressive investment.

\begin{figure}[!htb]
\captionsetup{skip=-0.5em}
\caption{Report-based Forecasts and Abnormal Stock Returns: Firm Characteristics} \label{fig:xs}
\begin{center}
	\includegraphics[width =\textwidth]{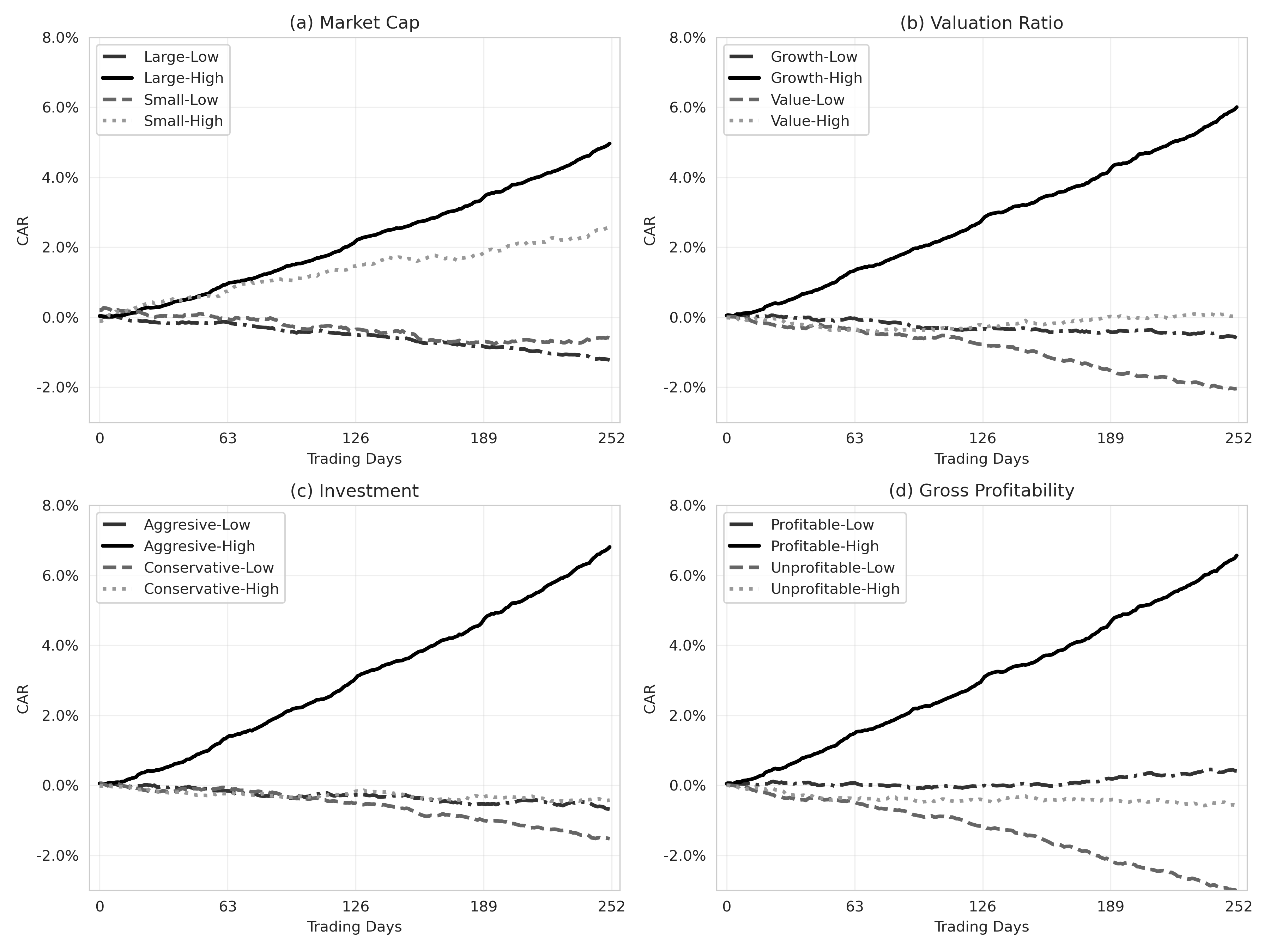}
\end{center}
\vspace{-0.2in}
{\footnotesize Note: This figure presents DGTW characteristic-adjusted buy-and-hold abnormal returns following research report releases, measured over a [0, +252] trading day window after each announcement. Reports are sorted based on two dimensions. First, they are categorized by the monthly median of the following characteristics at the end of the previous month: (a) market capitalization, (b) BM ratio, (c) investment, and (d) gross profitability. Second, within each category, reports are sorted into deciles based on their predicted 12-month returns. Returns are value-weighted within each calendar month using lagged market-capitalization weights and then equally averaged across months. The sample spans from 2005 to 2023.}
\end{figure}

Figure \ref{fig:xs} visualizes these conditional results by plotting DGTW-adjusted buy-and-hold abnormal returns over the 252-day window. The figure confirms the tabulated results, showing that predictive power is concentrated in large-cap, growth-oriented, high-investment, and profitable firms, where the long-short spreads reach approximately 6\%, while the corresponding strategies in small-cap, value, low-investment, and low-profitability firms show significantly smaller spreads. Figure \ref{fig:xs_ana}, in contrast, demonstrates that analyst characteristics have minimal impact on the signal's predictive performance. Across all four analyst dimensions—forecast accuracy, experience, broker size, and industry coverage—the long-short spreads are remarkably similar between the above-median and below-median groups, with all strategies generating spreads in the 4-7\% range. This uniform performance across analyst types reinforces the finding from Panel A of Table \ref{tab:char_stats} that the narrative signal's effectiveness is not driven by analyst traits or institutional resources, but rather captures genuine information content that transcends individual analyst characteristics.

I next examine the contemporaneous relationship between the report-narrative portfolio (RP) and returns on popular traded factors (Panel A), as well as the Fama-French 12 industry portfolios (Panel B). The findings in Table \ref{tab:corr} reveal a distinct factor exposure profile. Among traded factors, RP exhibits the strongest correlations with investment and value style proxies: correlations with CMA and \text{R\_IA} are $58.4\%$ and $59.1\%$, respectively, while the correlation with HML reaches $53.6\%$. These negative loadings are consistent with a pronounced growth and aggressive investment orientation. In contrast, correlations with size (SMB: 3.3\%), profitability (RMW: $0.5\%$, \text{R\_ROE}: $8.8\%$), momentum (MOM: $5.9\%$), and alternative size measures (\text{R\_ME}: $8.3\%$) remain modest. RP exhibits a moderate pro-cyclical profile, with positive correlations to MKT-RF ($24.7\%$) and earnings-growth \text{R\_EG} ($21.3\%$).

The industry analysis reinforces this growth-oriented pattern. RP displays the strongest correlations with Business Equipment (45.1\%), Consumer Durables (33.5\%), and Shops ($28.7\%$), followed by Manufacturing ($17.1\%$). Notably, correlations with defensive sectors are negligible, for example, Utilities ($1.0\%$), Money ($0.2\%$ ), and Nondurables ($0.5 \%$).

This concentration pattern reveals that the narrative-based strategy is systematically tilted toward large, profitable growth firms, particularly in technology sectors with aggressive investment profiles. The stark contrast between uniform performance across analyst characteristics versus pronounced variation across firm characteristics indicates that the textual analysis captures fundamental company information that is especially valuable for certain stock types, rather than analyst-specific expertise or institutional advantages.

\subsubsection{Contemporaneous News} \label{news}

Since analyst reports are often issued around major corporate announcements, a key question is whether the narrative signal simply repackages post-earnings-announcement drift (PEAD) documented by \citet{livnat2006comparing} rather than capturing unique textual information. If the predictive power stems from market underreaction to public earnings news, we would expect reports with higher earnings surprises and stronger immediate market reactions to generate higher narrative-based return forecasts.

To test this hypothesis, I examine the relationship between return prediction deciles and several key measures. Table \ref{tab:sentiment} presents the latest earnings surprise (SUE) prior to report releases and two-day DGTW characteristic-adjusted abnormal returns ($CAR_{[0,+1]}$) following report releases, which capture earnings news and immediate market reactions, respectively. If the PEAD mechanism drives the observed predictability, firms with higher SUE and $CAR_{[0,+1]}$ should systematically receive higher narrative-based return forecasts. 

The results contradict the PEAD hypothesis. While the model successfully predicts relative performance, with realized returns ($Ret_{12m}$) showing a significant 4.9\% high-low spread that aligns closely with earlier findings. The relationship with traditional PEAD drivers is either weak or contrary to expectations. Earnings surprises (SUE) exhibit significant variation across forecast deciles, with a -0.007\% high-low spread. The immediate market reaction ($CAR_{[0,+1]}$) also shows a significant negative relationship with narrative forecasts: reports generating the highest return predictions actually experience the weakest contemporaneous market reactions, with a high-low spread of -25.0\%. 

This negative correlation suggests that the narrative signal captures information orthogonal to immediate market reactions and earnings surprises. Rather than identifying stocks benefiting from delayed price adjustment to earnings news, the model appears to detect more subtle information embedded in analyst language that the market initially underappreciates or overreacts to in the short term.

The divergence between narrative forecasts and immediate market reactions extends systematically to analysts' own quantitative assessments. Table \ref{tab:sentiment} reveals that recommendation revisions ($REC_{REV}$), earnings forecast revisions ($EF_{REV}$), and target price revisions ($TP_{REV}$) all show modest negative correlations with narrative-based return predictions. This pattern indicates that the textual signal often contradicts analysts' explicit numerical guidance, suggesting the narrative content captures information that analysts either cannot or choose not to reflect fully in their formal recommendations.

Taken together, this conditional analysis reveals an important asymmetry in how narrative information is processed and incorporated by markets. The textual content appears most informative precisely when it provides a contrarian perspective. When immediate market reactions are muted, earnings surprises are modest, and analysts' quantitative revisions are conservative. This pattern, where the highest-return predictions coincide with the most negative contemporaneous signals, suggests markets temporarily overreact to short-term news while underweighting the long-term fundamentals embedded in analyst narratives. Such behavior is more consistent with gradual mispricing correction than with stable risk compensation, supporting the view that textual analysis captures information processing inefficiencies that persist during periods of negative sentiment.

Given the results, I next test two plausible explanations: that the narrative signal is a proxy for short-term sentiment, or that it profits from the reversal of an initial announcement-window overreaction. I develop four distinct sentiment-based strategies using: (i) the two-day DGTW-adjusted abnormal return around report dates, $CAR_{[0,+1]}$; (ii) recommendation revisions; (iii) headline tone; and (iv) body tone from a fine-tuned BERT classifier. Following the baseline methodology, I rank stocks by their trailing twelve-month average sentiment scores and assess subsequent one-month returns.

Table \ref{tab:sentport} presents the results, which provide little support for sentiment-based or overreaction-based explanations of the narrative signal's predictive power. A strategy based on short-term announcement returns ($CAR_{[0,+1]}$) fails to generate meaningful profits, producing a negative and statistically insignificant long-short alpha of -0.36\% per month (t = -1.57). This finding contradicts the simple announcement overreaction hypothesis.

The recommendation revision strategy shows the most promise among the alternatives, generating modest raw returns, but its risk-adjusted performance remains weak with an alpha of only 0.24\% per month (t = 1.56). The document-level sentiment strategies also fail to replicate the narrative signal's performance. Neither headline nor body tone exhibits a monotonic relationship with future returns, generating economically and statistically negligible long-short alphas. This suggests that simple sentiment scoring, even using sophisticated NLP techniques, cannot capture the predictive information embedded in the narrative analysis.

The systematic failure of these sentiment-based strategies to replicate the narrative signal's predictive power indicates that the model captures information beyond simple market sentiment or tone. Instead, the results suggest that the narrative analysis identifies more subtle linguistic patterns and contextual information that traditional sentiment measures fail to detect.

\subsection{What Content Matters?} \label{shap}

I now investigate the sources of investment value with the Shapley value decomposition framework described in Section \ref{topic_modeling}. For each meta-category, Table \ref{tab:shap} reports its share of report text and its Shapley contribution to portfolio performance. Financial Analysis and Company \& Industry Overview account for the largest shares of content, representing 36.56\% and 28.53\% of total coverage, respectively. Strategic Outlook and Risk \& Governance form a second tier at 15.13\% and 14.14\%, and Additional Content accounts for the remaining 5.63\%.

The Shapley value decomposition reveals a different pattern. Strategic Outlook emerges as the most valuable category, accounting for 41.34\% and 31.43\% of the portfolio's Sharpe ratio and returns, respectively, despite not being the most extensively discussed component. Company \& Industry Overview maintains proportional importance, contributing 27.61\% to the Sharpe ratio and 26.92\% to returns. Risk and Governance shows outsized impact on returns relative to its discussion frequency, accounting for 11.21\% of Sharpe ratio value and 21.36\% of returns. Additional Content has minimal impact on performance. 

Despite constituting the most extensively covered category (36.56\% of content), Financial Analysis discussions contribute only to 16.39\% of the Sharpe ratio and 19.53\% of returns. This pattern is consistent with standardized financial data being incorporated into prices more quickly, leaving less scope for subsequent drift. In turn, forward-looking strategic assessments and industry context appear to diffuse more slowly, generating larger abnormal returns. Overall, Table \ref{tab:shap} suggests that the investment value in analyst reports is concentrated in strategic, forward-looking content rather than financial analysis.

Table \ref{tab:soshap} further decomposes the value of the Strategic Outlook category along three dimensions—timeframe, sentiment, and analytical focus. For each dimension, I classify sentences using a structured prompt that categorizes timeframe (long-term, short-term, or both), sentiment (negative, neutral, or positive), and focus (risk versus fundamental analysis). I use ChatGPT-4o to produce initial labels with the prompt below and then train three BERT classifiers on these labels for corpus-scale inference.

\vspace{0.2in}
\noindent
\begin{tabular}{|p{16cm}|}
\hline
\textbf{Prompt:} 

Please read the following sentence from a sell-side analyst report (investment thesis section) carefully and classify it into three numeric labels:

\textbf{Timeframe}:
   1 = Long-term (multi-quarter or multi-year outlook);
   2 = Short-term (next quarter or near-term);
   3 = Both (if it clearly addresses both short- and long-term).

\textbf{Sentiment}:
   1 = Positive potential;
   0 = Neutral;
   1 = Negative/Bearish.

\textbf{Focus}:
   1 = Fundamentals (e.g., belief in growth strategy, industry tailwinds, earnings);
   0 = Cautious risk (e.g., warnings, near-term headwinds, legal/regulatory risk).

\textbf{Important note:} If the sentence mentions both short-term risk and a fundamental (longer-term) driver, prioritize the fundamental aspect and mark the third label as \textbf{1} (Fundamentals).

\textbf{Output format:} Only output the three numbers in the format: \texttt{[X, Y, Z]}.

Here is the sentence: \{sentence\}
\\
\hline
\end{tabular}
\vspace{0.2in}

The decomposition of performance along these dimensions reveals a clear hierarchy of value. First, long-term discussions are the primary driver, accounting for approximately half of the portfolio's Sharpe ratio (49.51\%) and returns (50.20\%). Second, positive sentiment is paramount, contributing nearly 50\% to performance, with neutral and negative sentiments making up the remainder in roughly equal measure. Finally, the analysis of focus shows that discussions of fundamentals are dominantly important, driving 87.00\% of the Sharpe ratio, while risk-focused commentary contributes comparatively little.

Together, these patterns indicate that the predictive power of Strategic Outlook stems primarily from long-term, positive assessments of fundamentals rather than from short-term commentary or risk warnings, suggesting that the market is gradually incorporating analysts' long-term fundamental assessments.

A natural follow-up is whether isolating the most value-relevant content yields cleaner signals. In Table \ref{tab:catdecile}, I report the value-weighted decile portfolio performance sorted by category-specific content. I focus on four categories, excluding additional content that primarily consists of boilerplate and residual text.

Strategic Outlook content generates the strongest portfolio performance, with the long-short strategy delivering a mean return of 1.41\% per month and a Sharpe ratio of 0.93. Both legs contribute, and the top decile earns 1.87\% per month with a significant alpha (0.72, t = 2.85). This superiority aligns with the Shapley results that identify Strategic Outlook as the primary source of value.

By comparison, Company \& Industry Overview, Financial Analysis, and Risk \& Governance produce more modest long--short returns (0.52\%--0.58\% per month) and Sharpe ratios (0.37--0.41). While the highest deciles of these categories show some predictive power, particularly Risk \& Governance with an alpha of 0.32 (t = 1.97), their overall performance is substantially weaker than that of Strategic Outlook. This pattern is again consistent with the Shapley value decomposition, validating the effectiveness of the strategy.

The disparity in univariate portfolio performances provides additional support for a mispricing interpretation. If the narrative factor captures systematic risk, one would expect Risk \& Governance content to generate the strongest returns, as this category explicitly discusses risk factors and uncertainties. Instead, the forward-looking Strategic Outlook content dominates performance, suggesting that markets are slow to process complex assessments about future prospects rather than compensating investors for bearing systematic risk. The weak performance of Financial Analysis content, despite its prevalence in reports, further supports the view that standardized quantitative information is efficiently priced, while qualitative forward-looking judgments create temporary overreaction that corrects over time.

I next examine the incremental contribution of Strategic Outlook. As reported in Table \ref{tab:IIV_SO}, the Strategic Outlook portfolio delivers information ratios of 0.96--1.41 when evaluated against both the 94 fundamental predictors of \citet{gu2020empirical} and the 18 analyst-based factors of \citet{ChenZimmermann2021}. Table \ref{tab:lab_so} confirms that the portfolio performance is not driven by lookahead bias. Across alternative language model representations, the Strategic Outlook strategy attains a post-knowledge cutoff Sharpe ratio between 0.65 and 2.37. Taken together, the two tables reinforce the central interpretation: long-horizon investment value in analyst narratives is concentrated in forward-looking strategic assessments, and that content alone is sufficient to generate robust, high-efficiency portfolios.

\section{Conclusion}
\label{sec:conclu}

I document novel investment value embedded in the textual narratives of analyst reports. By representing these narratives as high-dimensional LLM embeddings, I show they predict future stock returns. A portfolio strategy based on these signals delivers economically and statistically significant alpha that is incremental to analysts' quantitative outputs and a vast set of established factors and characteristic-based predictors.

The report narratives' predictive power is not uniform.  It is more pronounced for growth firms with aggressive investment. Factor regressions show consistent tilts (negative value and investment), yet the alpha remains statistically and economically meaningful. Predictability is more favorable when contemporaneous news is adverse. For example, following negative earnings surprises or recommendation downgrades. However, coarse report sentiment measures and announcement‑window reactions do not reproduce the effect.

Using a Shapley value decomposition built from sentence‑level topics aggregated to five meta‑categories, I attribute performance to economically interpretable content. Strategic outlook dominates, with the largest contributions from investment thesis and valuation language that emphasize longer‑run fundamentals. A strategy based solely on this strategic outlook content exceeds the full model's alpha, indicating that the incremental information is highly concentrated in this forward-looking section.

These findings provide direct evidence that analyst narratives embed forward‑looking assessments that the market incorporates gradually. The evidence collectively supports a mispricing interpretation of analyst narrative value. The concentration of predictive power in forward-looking strategic content, combined with stronger effects after adverse news, suggests markets are slow to process complex textual information about long-term fundamentals. Methodologically, the paper shows how modern language models can recover and attribute economic content from unstructured text without sacrificing interpretability. Substantively, these evidences complement \citet{cao2024man} by showing that generative AI serves to augment, rather than replace, human expertise. Leveraging the judgments of financial analysts, these tools allow the rich information embedded in qualitative research to be measured, tested, and deployed in a disciplined way.

\setlength{\bibsep}{4pt}
\bibliographystyle{jf}
\newpage
\bibliography{references}

\begin{thebibliography}{37}
\expandafter\ifx\csname natexlab\endcsname\relax\def\natexlab#1{#1}\fi

\bibitem[Alt{\i}nk{\i}l{\i}{\c{c}} et~al.(2016)Alt{\i}nk{\i}l{\i}{\c{c}}, Hansen, and Ye]{altinkilicc2016can}
Alt{\i}nk{\i}l{\i}{\c{c}}, Oya, Robert~S Hansen, and Liyu Ye, 2016, Can analysts pick stocks for the long-run?, {\em Journal of Financial Economics\/} 119, 371--398.

\bibitem[Birru et~al.(2024)Birru, Gokkaya, Liu, and Stulz]{birru2024analyst}
Birru, Justin, Sinan Gokkaya, Xi~Liu, and Ren{\'e} Stulz, 2024, Are analyst “top picks” informative?, {\em The Review of Financial Studies\/} 37, 1538--1583.

\bibitem[Birru et~al.(2022)Birru, Gokkaya, Liu, and Stulz]{birru2022analyst}
Birru, Justin, Sinan Gokkaya, Xi~Liu, and Ren{\'e}~M Stulz, 2022, Are analyst short-term trade ideas valuable?, {\em The Journal of Finance\/} 77, 1829--1875.

\bibitem[Bordalo et~al.(2019)Bordalo, Gennaioli, Porta, and Shleifer]{bordalo2019diagnostic}
Bordalo, Pedro, Nicola Gennaioli, Rafael~La Porta, and Andrei Shleifer, 2019, Diagnostic expectations and stock returns, {\em The Journal of Finance\/} 74, 2839--2874.

\bibitem[Brav and Lehavy(2003)]{brav2003empirical}
Brav, Alon, and Reuven Lehavy, 2003, An empirical analysis of analysts' target prices: Short-term informativeness and long-term dynamics, {\em The Journal of Finance\/} 58, 1933--1967.

\bibitem[Bybee et~al.(2024)Bybee, Kelly, Manela, and Xiu]{bybee2024business}
Bybee, Leland, Bryan Kelly, Asaf Manela, and Dacheng Xiu, 2024, Business news and business cycles, {\em The Journal of Finance\/} 79, 3105--3147.

\bibitem[Bybee et~al.(2023)Bybee, Kelly, and Su]{bybee2023narrative}
Bybee, Leland, Bryan Kelly, and Yinan Su, 2023, Narrative asset pricing: Interpretable systematic risk factors from news text, {\em The Review of Financial Studies\/} 36, 4759--4787.

\bibitem[Cao et~al.(2024)Cao, Jiang, Wang, and Yang]{cao2024man}
Cao, Sean, Wei Jiang, Junbo Wang, and Baozhong Yang, 2024, From man vs. machine to man+ machine: The art and ai of stock analyses, {\em Journal of Financial Economics\/} 160, 103910.

\bibitem[Chen and Zimmermann(2022)]{ChenZimmermann2021}
Chen, Andrew~Y., and Tom Zimmermann, 2022, Open source cross-sectional asset pricing, {\em Critical Finance Review\/} 27, 207--264.

\bibitem[Chen et~al.(2023{\natexlab{a}})Chen, Hwang, and Peng]{chen2023revisiting}
Chen, Hailiang, Byoung-Hyoun Hwang, and Zhuozhen Peng, 2023{\natexlab{a}}, Revisiting the cross-section of expected stock returns: Evidence from a textual analysis of buy recommendations, {\em Nanyang Business School Research Paper\/} .

\bibitem[Chen et~al.(2023{\natexlab{b}})Chen, Pelger, and Zhu]{chen2023deep}
Chen, Luyang, Markus Pelger, and Jason Zhu, 2023{\natexlab{b}}, Deep learning in asset pricing, {\em Management Science\/} .

\bibitem[Chen et~al.(2022)Chen, Kelly, and Xiu]{chen2022expected}
Chen, Yifei, Bryan~T Kelly, and Dacheng Xiu, 2022, Expected returns and large language models, {\em Available at SSRN 4416687\/} .

\bibitem[Christophe et~al.(2010)Christophe, Ferri, and Hsieh]{christophe2010informed}
Christophe, Stephen~E, Michael~G Ferri, and Jim Hsieh, 2010, Informed trading before analyst downgrades: Evidence from short sellers, {\em Journal of Financial Economics\/} 95, 85--106.

\bibitem[Daniel et~al.(1997)Daniel, Grinblatt, Titman, and Wermers]{daniel1997measuring}
Daniel, Kent, Mark Grinblatt, Sheridan Titman, and Russ Wermers, 1997, Measuring mutual fund performance with characteristic-based benchmarks, {\em The Journal of Finance\/} 52, 1035--1058.

\bibitem[Daniel et~al.(2020)Daniel, Hirshleifer, and Sun]{daniel2020short}
Daniel, Kent, David Hirshleifer, and Lin Sun, 2020, Short-and long-horizon behavioral factors, {\em The Review of Financial Studies\/} 33, 1673--1736.

\bibitem[DeMiguel et~al.(2009)DeMiguel, Garlappi, and Uppal]{demiguel2009optimal}
DeMiguel, Victor, Lorenzo Garlappi, and Raman Uppal, 2009, Optimal versus naive diversification: How inefficient is the 1/n portfolio strategy?, {\em The Review of Financial Studies\/} 22, 1915--1953.

\bibitem[DeMiguel et~al.(2020)DeMiguel, Martin-Utrera, Nogales, and Uppal]{demiguel2020transaction}
DeMiguel, Victor, Alberto Martin-Utrera, Francisco~J Nogales, and Raman Uppal, 2020, A transaction-cost perspective on the multitude of firm characteristics, {\em The Review of Financial Studies\/} 33, 2180--2222.

\bibitem[Fama and French(2015)]{fama2015five}
Fama, Eugene~F, and Kenneth~R French, 2015, A five-factor asset pricing model, {\em Journal of Financial Economics\/} 116, 1--22.

\bibitem[Fama and French(2018)]{fama2018choosing}
Fama, Eugene~F, and Kenneth~R French, 2018, Choosing factors, {\em Journal of Financial Economics\/} 128, 234--252.

\bibitem[Farago et~al.(2023)Farago, Hjalmarsson, and Zeng]{farago2023analysts}
Farago, Adam, Erik Hjalmarsson, and Ming Zeng, 2023, Analysts are good at ranking stocks, {\em Available at SSRN 4495163\/} .

\bibitem[Green(2006)]{green2006value}
Green, T~Clifton, 2006, The value of client access to analyst recommendations, {\em Journal of Financial and Quantitative Analysis\/} 41, 1--24.

\bibitem[Gu et~al.(2020)Gu, Kelly, and Xiu]{gu2020empirical}
Gu, Shihao, Bryan Kelly, and Dacheng Xiu, 2020, Empirical asset pricing via machine learning, {\em The Review of Financial Studies\/} 33, 2223--2273.

\bibitem[He et~al.(2025)He, Lv, Manela, and Wu]{he2025chronologically}
He, Songrun, Linying Lv, Asaf Manela, and Jimmy Wu, 2025, Chronologically consistent large language models, {\em arXiv preprint arXiv:2502.21206\/} .

\bibitem[Hirshleifer et~al.(forthcoming)Hirshleifer, Mai, and Pukthuanthong]{hirshleifer2023war}
Hirshleifer, David, Dat Mai, and Kuntara Pukthuanthong, forthcoming, War discourse and the cross section of expected stock returns, {\em The Journal of Finance\/} .

\bibitem[Hodrick(1992)]{hodrick1992dividend}
Hodrick, Robert~J, 1992, Dividend yields and expected stock returns: Alternative procedures for inference and measurement, {\em The Review of Financial Studies\/} 5, 357--386.

\bibitem[Hou et~al.(2015)Hou, Xue, and Zhang]{hou2015digesting}
Hou, Kewei, Chen Xue, and Lu~Zhang, 2015, Digesting anomalies: An investment approach, {\em The Review of Financial Studies\/} 28, 650--705.

\bibitem[Jha et~al.(2025)Jha, Liu, and Manela]{jha2025does}
Jha, Manish, Hongyi Liu, and Asaf Manela, 2025, Does finance benefit society? a language embedding approach, {\em The Review of Financial Studies\/} .

\bibitem[Li et~al.(2024)Li, Mai, Shen, Yang, and Zhang]{li2024dissecting}
Li, Kai, Feng Mai, Rui Shen, Chelsea Yang, and Tengfei Zhang, 2024, Dissecting corporate culture using generative ai--insights from analyst reports, {\em Available at SSRN 4558295\/} .

\bibitem[Livnat and Mendenhall(2006)]{livnat2006comparing}
Livnat, Joshua, and Richard~R Mendenhall, 2006, Comparing the post--earnings announcement drift for surprises calculated from analyst and time series forecasts, {\em Journal of Accounting Research\/} 44, 177--205.

\bibitem[Lv(2024)]{lv2024value}
Lv, Linying, 2024, The value of information from sell-side analysts, {\em arXiv preprint arXiv:2411.13813\/} .

\bibitem[Lys and Sohn(1990)]{lys1990association}
Lys, Thomas, and Sungkyu Sohn, 1990, The association between revisions of financial analysts' earnings forecasts and security-price changes, {\em Journal of Accounting and Economics\/} 13, 341--363.

\bibitem[Malmendier and Shanthikumar(2014)]{malmendier2014security}
Malmendier, Ulrike, and Devin Shanthikumar, 2014, Do security analysts speak in two tongues?, {\em The Review of Financial Studies\/} 27, 1287--1322.

\bibitem[Manela and Moreira(2017)]{manela2017news}
Manela, Asaf, and Alan Moreira, 2017, News implied volatility and disaster concerns, {\em Journal of Financial Economics\/} 123, 137--162.

\bibitem[Meursault et~al.(2023)Meursault, Liang, Routledge, and Scanlon]{meursault2023pead}
Meursault, Vitaly, Pierre~Jinghong Liang, Bryan~R Routledge, and Madeline~Marco Scanlon, 2023, Pead. txt: Post-earnings-announcement drift using text, {\em Journal of Financial and Quantitative Analysis\/} 58, 2299--2326.

\bibitem[Sarkar and Vafa(2024)]{sarkar2024lookahead}
Sarkar, Suproteem~K, and Keyon Vafa, 2024, Lookahead bias in pretrained language models, {\em Available at SSRN\/} .

\bibitem[Siano(2025)]{siano2025news}
Siano, Federico, 2025, The news in earnings announcement disclosures: Capturing word context using llm methods, {\em Management Science\/} .

\bibitem[Tetlock(2007)]{tetlock2007giving}
Tetlock, Paul~C, 2007, Giving content to investor sentiment: The role of media in the stock market, {\em The Journal of Finance\/} 62, 1139--1168.

\end{thebibliography}

\newpage

\newpage

\newpage


\newpage
\begin{table}[!htb]
    \caption{Summary Statistics of Analyst Reports}\label{tab:sum1}
    {\footnotesize This table presents summary statistics for analyst reports covering S\&P 1500 firms from 2000 to 2023. For each year, it reports the number of research reports, distinct brokerage firms, unique sell-side analysts, and the average report characteristics (number of pages, sentences, and tokens per report).}
    \footnotesize
    \begin{center}
    \begin{tabularx}{\textwidth}{YYYYYYY}
    \toprule
    Year & Reports & Brokers & Analysts & Pages & Sentences & Tokens \\
    \midrule
    2000 & 14224 & 49 & 555 & 5 & 56 & 1244 \\
    2001 & 22309 & 50 & 639 & 5 & 57 & 1256 \\
    2002 & 27555 & 55 & 791 & 5 & 57 & 1172 \\
    2003 & 28323 & 67 & 859 & 6 & 64 & 1257 \\
    2004 & 33431 & 70 & 969 & 6 & 66 & 1210 \\
    2005 & 39365 & 71 & 967 & 6 & 64 & 1167 \\
    2006 & 42347 & 70 & 951 & 6 & 64 & 1147 \\
    2007 & 43742 & 69 & 970 & 7 & 69 & 1217 \\
    2008 & 44676 & 72 & 1070 & 7 & 73 & 1288 \\
    2009 & 39425 & 83 & 1104 & 7 & 73 & 1281 \\
    2010 & 28998 & 86 & 1028 & 7 & 75 & 1312 \\
    2011 & 59367 & 86 & 1279 & 7 & 76 & 1263 \\
    2012 & 66576 & 85 & 1269 & 8 & 76 & 1205 \\
    2013 & 66960 & 78 & 1202 & 7 & 72 & 1143 \\
    2014 & 65748 & 73 & 1176 & 7 & 69 & 1086 \\
    2015 & 66699 & 75 & 1137 & 8 & 71 & 1138 \\
    2016 & 66891 & 71 & 1117 & 8 & 74 & 1170 \\
    2017 & 66846 & 67 & 1010 & 9 & 78 & 1234 \\
    2018 & 64310 & 62 & 930 & 9 & 82 & 1253 \\
    2019 & 64271 & 60 & 972 & 9 & 83 & 1287 \\
    2020 & 64567 & 63 & 964 & 10 & 85 & 1297 \\
    2021 & 56793 & 61 & 971 & 9 & 87 & 1300 \\
    2022 & 57836 & 60 & 981 & 10 & 92 & 1399 \\
    2023 & 63071 & 59 & 972 & 10 & 91 & 1353 \\
    \bottomrule
    \end{tabularx}
    \end{center}
\end{table}

\newpage
\begin{landscape}
\begin{table}[!htb]
\caption{Return Predictability}
\label{tab:predictability}
    {\footnotesize This table reports the results from regressions of the specification $RET_{i,12m}= \alpha_{t} + \beta^{\prime} x_{i, t}+ \varepsilon_{i, t+12}$, where $RET_{i,12m}$ is the future realized return of stock $i$. $x_{i, t}$ represents analyst information from each report. $\widehat{RET}_{12m}$ is the predicted next 12 months' stock return from ridge regressions. $REC_{REV}$ denotes recommendation revision, calculated as the current report's recommendation minus the last recommendation in I/B/E/S issued by the same analyst for the same stock. $EF_{REV}$ refers to earnings forecast revision, calculated as the current report's EPS forecast minus the last EPS forecast in I/B/E/S issued by the same analyst for the same stock, scaled by the stock price 50 days before the report release. $TP_{REV}$ represents the target price revision, calculated as the current report's target price minus the last target price in I/B/E/S issued by the same analyst for the same stock, scaled by the stock price 50 days before the report release. $Tone$ assesses the report content sentiment. I include analyst-year-month fixed effect and industry-year-month fixed effect. Standard errors are two-way clustered by firm and year-month. The t-statistics are shown in parentheses, with ***, **, and * indicating statistical significance at the $1 \%, 5 \%$, and $10 \%$ level, respectively. The sample spans from 2005 to 2023.}
\footnotesize
    \begin{center}
\begin{tabularx}{1.4\textwidth}{lYYYYYYY}
\toprule
\multicolumn{8}{c}{\textbf{Panel A: Forecast vs.\ Realized Returns Across Horizons}} \\
\midrule
                    &\multicolumn{1}{c}{$RET_{1m}$}&\multicolumn{1}{c}{$RET_{3m}$}&\multicolumn{1}{c}{$RET_{6m}$}&\multicolumn{1}{c}{$RET_{9m}$}&\multicolumn{1}{c}{$RET_{12m}$}&\multicolumn{1}{c}{$RET_{18m}$}&\multicolumn{1}{c}{$RET_{24m}$}\\
\midrule
$\widehat{RET}_{12m}$ &       0.003\sym{*}  &       0.012\sym{***}&       0.026\sym{***}&       0.043\sym{***}&       0.060\sym{***}&       0.091\sym{***}&       0.114\sym{***}\\
                    &      (1.85)         &      (3.64)         &      (4.85)         &      (4.75)         &      (4.37)         &      (5.17)         &      (5.24)         \\
\midrule
Ana $\times$ YM FE & Yes & Yes & Yes & Yes & Yes & Yes & Yes \\
Ind $\times$ YM FE & Yes & Yes & Yes & Yes & Yes & Yes & Yes \\
Adjusted \(R^{2}\)  &       0.534         &       0.578         &       0.574         &       0.562         &       0.491         &       0.524         &       0.528         \\
$N$       &      488,151         &      488,151         &      488,151         &      488,151         &      488,151         &      473,830         &      459,623         \\
\midrule
\end{tabularx}
\begin{tabularx}{1.4\textwidth}{lYYYYYYY}
\multicolumn{8}{c}{\textbf{Panel B: Numerical versus Textual Signals}} \\
\midrule
                    &\multicolumn{1}{c}{$RET_{12m}$}&\multicolumn{1}{c}{$RET_{12m}$}&\multicolumn{1}{c}{$RET_{12m}$}&\multicolumn{1}{c}{$RET_{12m}$}&\multicolumn{1}{c}{$RET_{12m}$}&\multicolumn{1}{c}{$RET_{12m}$}&\multicolumn{1}{c}{$RET_{12m}$}\\
                    &\multicolumn{1}{c}{(1)}&\multicolumn{1}{c}{(2)}&\multicolumn{1}{c}{(3)}&\multicolumn{1}{c}{(4)}&\multicolumn{1}{c}{(5)}&\multicolumn{1}{c}{(6)}&\multicolumn{1}{c}{(7)}\\
\midrule
$\widehat{RET}_{12m}$ &                     &                     &                     &                     &       0.087\sym{***}&       0.100\sym{***}&       0.070\sym{***}\\
                    &                     &                     &                     &                     &      (3.57)         &      (3.34)         &      (4.87)         \\
$REC_{REV}$        &      -0.001         &                     &                     &                     &                     &      -0.005         &      -0.001         \\
                    &     (-0.18)         &                     &                     &                     &                     &     (-0.59)         &     (-0.07)         \\
$EF_{REV}$         &                     &      -0.351         &                     &                     &                     &      -0.271         &       0.047         \\
                    &                     &     (-0.90)         &                     &                     &                     &     (-0.59)         &      (0.11)         \\
$TP_{REV}$         &                     &                     &       0.006         &                     &                     &       0.039         &       0.022         \\
                    &                     &                     &      (0.25)         &                     &                     &      (1.17)         &      (1.09)         \\
Tone           &                     &                     &                     &      -0.009         &                     &      -0.008         &      -0.013         \\
                    &                     &                     &                     &     (-1.22)         &                     &     (-0.89)         &     (-1.61)         \\
\midrule
YM FE & No & No & No & No & Yes &Yes & No \\
Ana $\times$ YM FE & Yes & Yes & Yes & Yes & No & No & Yes \\
Ind $\times$ YM FE & Yes & Yes & Yes & Yes & No & No & Yes \\
Adjusted \(R^{2}\)  &       0.489         &       0.573         &       0.474         &       0.490         &       0.162         &       0.165         &       0.579         \\
$N$ &      480,523         &      344,742         &      341,049         &      487,304         &      837,233         &      286,129         &      247,676         \\
\bottomrule
\end{tabularx}
    \end{center}
\end{table}
\end{landscape}

\newpage
\begin{table}[!htb]
\caption{Report-based Forecasts and One-year Abnormal Stock Returns}
\label{tab:ls_stat}
\footnotesize
{\footnotesize This table reports mean DGTW characteristic-adjusted buy-and-hold abnormal returns (\%) over 252 trading days for value-weighted portfolios sorted on report-based return predictions. ``High'' and ``Low'' refer to the top and bottom deciles of predicted returns, respectively. ``High--Low'' represents the return spread between ``High'' and ``Low'' deciles. Results are shown for six different language models: LLaMA3-8B, ChronoGPT$_{1999}$, ChronoGPT$_{2024}$, BERT, RoBERTa, and LLaMA2-13B. \citet{hodrick1992dividend} t-statistics are reported in parentheses. ***, **, and * indicate significance at the 1\%, 5\%, and 10\% levels, respectively. The sample spans from 2005 to 2023.}
\begin{center}
\begin{tabularx}{\textwidth}{@{\hskip\tabcolsep\extracolsep\fill}lYYY}
\toprule
Model    & \multicolumn{1}{c}{High} & \multicolumn{1}{c}{Low} & \multicolumn{1}{c}{High--Low} \\
\midrule
LLaMA3         & 0.047\sym{***}  & -0.012  & 0.059\sym{***} \\
               & (3.99)          & (-1.31) & (3.46) \\
$\text{ChronoGPT}_{1999}$ & 0.035\sym{***} & -0.005 & 0.040\sym{***} \\
               & (3.36)          & (-0.74) & (2.89) \\
$\text{ChronoGPT}_{2024}$ & 0.033\sym{***} & -0.007 & 0.040\sym{***} \\
               & (3.40)          & (-0.90) & (3.02) \\
BERT           & 0.018\sym{**}   & -0.002  & 0.021\sym{**} \\
               & (2.58)          & (-0.33) & (2.21) \\
RoBERTa        & 0.016\sym{**}   & -0.003  & 0.019\sym{*} \\
               & (2.37)          & (-0.37) & (1.90) \\
LLaMA2         & 0.039\sym{***}  & -0.011  & 0.050\sym{***} \\
               & (3.44)          & (-1.21) & (3.10) \\
\bottomrule
\end{tabularx}
\end{center}
\end{table}

\newpage
\begin{table}[!htb]
\footnotesize
\caption{Portfolio Statistics}
\label{tab:decile}
    {\footnotesize This table presents the performance of value-weighted decile portfolios sorted by ridge models' return forecasts, which are based on the past \{LB\} months of analyst report information. The models predict 12-month-ahead stock returns. For each portfolio, I report several performance measures: excess return mean and standard deviation, Sharpe ratio, $\alpha$ relative to \citet{fama2018choosing} six factors, and the corresponding t-statistics for $\alpha$. Portfolios are rebalanced monthly using the average of the most recent 12 months of out-of-sample predictions. The `H-L' row represents a long-short strategy that takes a long position in the highest decile (High) and a short position in the lowest decile (Low). The sample period extends from January 2005 to June 2024.}
    \begin{center}
    \begin{tabularx}{\textwidth}{@{\hskip\tabcolsep\extracolsep\fill}l*{10}r}
    \toprule
    & \multicolumn{5}{c}{LB = 9 months} & \multicolumn{5}{c}{LB = 12 months} \\
    \cmidrule{2-6} \cmidrule{7-11}
     & Mean & SD & SR & $\alpha$ & $t_\alpha$  & Mean & SD & SR & $\alpha$ & $t_\alpha$ \\
    \midrule
    Low (L) & 0.53 & 4.68 & 0.39 & -0.09 & -0.45 & 0.52 & 4.90 & 0.37 & -0.13 & -0.68 \\
    2 & 0.74 & 4.76 & 0.54 & 0.06 & 0.56 & 0.63 & 4.53 & 0.48 & -0.05 & -0.44 \\
    3 & 0.65 & 4.34 & 0.52 & -0.11 & -1.38 & 0.66 & 4.53 & 0.50 & -0.15 & -1.56 \\
    4 & 0.71 & 4.54 & 0.54 & -0.05 & -0.53 & 0.77 & 4.45 & 0.60 & 0.04 & 0.39 \\
    5 & 0.75 & 4.71 & 0.55 & -0.05 & -0.60 & 0.75 & 4.53 & 0.58 & -0.04 & -0.55 \\
    6 & 0.72 & 4.53 & 0.55 & 0.00 & 0.02 & 0.84 & 4.76 & 0.61 & 0.07 & 0.62 \\
    7 & 0.81 & 4.89 & 0.57 & 0.04 & 0.30 & 0.73 & 4.75 & 0.53 & -0.06 & -0.58 \\
    8 & 0.95 & 4.68 & 0.70 & 0.13 & 1.61 & 1.07 & 4.83 & 0.76 & 0.25 & 2.07 \\
    9 & 0.94 & 5.32 & 0.61 & 0.06 & 0.34 & 0.78 & 5.32 & 0.51 & -0.08 & -0.61 \\
    High (H) & 1.52 & 6.38 & 0.82 & 0.50 & 2.66 & 1.56 & 6.29 & 0.86 & 0.55 & 2.73 \\
    H-L & 0.98 & 5.20 & 0.66 & 0.58 & 2.48 & 1.04 & 5.21 & 0.69 & 0.68 & 2.64 \\
    \end{tabularx}
    \vspace{0.3in}
    \begin{tabularx}{\textwidth}{@{\hskip\tabcolsep\extracolsep\fill}l*{10}r}
    \toprule
    & \multicolumn{5}{c}{LB = 18 months} & \multicolumn{5}{c}{LB = 24 months} \\
    \cmidrule{2-6} \cmidrule{7-11}
     & Mean & SD & SR & $\alpha$ & $t_\alpha$  & Mean & SD & SR & $\alpha$ & $t_\alpha$ \\
    \midrule
    Low (L) & 0.56 & 4.93 & 0.40 & -0.12 & -0.54 & 0.53 & 4.98 & 0.37 & -0.16 & -0.73 \\
    2 & 0.56 & 4.56 & 0.43 & -0.11 & -0.99 & 0.49 & 4.62 & 0.37 & -0.21 & -1.98 \\
    3 & 0.70 & 4.42 & 0.55 & -0.05 & -0.74 & 0.71 & 4.44 & 0.55 & -0.01 & -0.08 \\
    4 & 0.69 & 4.53 & 0.53 & -0.07 & -0.84 & 0.76 & 4.28 & 0.62 & 0.03 & 0.30 \\
    5 & 0.77 & 4.43 & 0.60 & 0.04 & 0.36 & 0.58 & 4.75 & 0.42 & -0.19 & -1.36 \\
    6 & 0.84 & 4.69 & 0.62 & -0.00 & -0.06 & 0.97 & 4.47 & 0.75 & 0.14 & 1.46 \\
    7 & 0.69 & 4.63 & 0.52 & -0.09 & -0.63 & 0.78 & 4.58 & 0.59 & -0.02 & -0.13 \\
    8 & 1.09 & 4.82 & 0.79 & 0.26 & 2.02 & 0.92 & 5.11 & 0.63 & 0.10 & 0.97 \\
    9 & 1.12 & 5.56 & 0.70 & 0.22 & 1.42 & 0.90 & 5.17 & 0.60 & -0.00 & -0.04 \\
    High (H) & 1.43 & 6.21 & 0.80 & 0.40 & 2.20 & 1.70 & 6.48 & 0.91 & 0.63 & 2.67 \\
    H-L & 0.87 & 4.84 & 0.62 & 0.52 & 1.99 & 1.16 & 5.13 & 0.79 & 0.79 & 2.48 \\
    \bottomrule
    \end{tabularx}
    \end{center}
\end{table}

\newpage
\begin{table}[!htb]
\caption{Incremental Investment Value}
\footnotesize
\label{tab:IIV}
    {\footnotesize This table examines the incremental investment value of analyst reports compared to existing factors. The analysis includes: full report content (RP), 18 analyst-based factors from \cite{ChenZimmermann2021} (ANA), and 92 fundamental-based factors from \cite{gu2020empirical} (ANOM). For each strategy and combination, I report mean returns, Sharpe ratio (SR), and information ratio (IR). Newey–West (HAC) t‑statistics with 12 lags for the average returns are shown in parentheses. I also report the alphas of long-short portfolios relative to four benchmark factor models: \citet{fama2015five} five factors ($\alpha_{F5}$), \citet{fama2018choosing} six factors ($\alpha_{F6}$), \citet{hou2015digesting} factors ($\alpha_{HXZ}$), and \citet{daniel2020short} factors ($\alpha_{DHS}$). The sample for traded factors is from January 2005 to December 2023.}
    \begin{center}
\begin{tabularx}{\textwidth}{lYYYYYYY}
\toprule
 & Mean & SR & $\alpha_{F5}$ & $\alpha_{F6}$ & $\alpha_{HXZ}$ & $\alpha_{DHS}$ & IR \\
\midrule
\multicolumn{8}{l}{Panel A: Factor Performances} \\
\midrule
RP & 1.07 & 0.71 & 0.73 & 0.75 & 0.86 & 1.21 & - \\
 & (3.07) & (3.04) & (2.73) & (2.86) & (3.52) & (3.90) & - \\
ANA & 0.27 & 0.48 & 0.28 & 0.25 & 0.22 & 0.16 & - \\
 & (2.08) & (2.07) & (2.80) & (3.01) & (2.08) & (1.26) & - \\
ANOM & 1.34 & 1.03 & 1.40 & 1.31 & 1.28 & 1.15 & - \\
 & (4.47) & (4.38) & (5.05) & (4.65) & (4.43) & (3.84) & - \\
ANA + ANOM & 0.80 & 1.02 & 0.84 & 0.78 & 0.75 & 0.66 & - \\
 & (4.44) & (4.35) & (5.30) & (5.16) & (4.36) & (3.58) & - \\
\midrule
\multicolumn{8}{l}{Panel B: Report versus Factors} \\
\midrule
RP + ANA & 0.67 & 0.90 & 0.51 & 0.50 & 0.54 & 0.68 & 0.73 \\
 & (3.91) & (3.85) & (3.94) & (3.92) & (4.81) & (5.34) & (2.81) \\
RP + ANOM & 1.21 & 1.55 & 1.06 & 1.03 & 1.07 & 1.18 & 1.16 \\
 & (6.75) & (6.44) & (7.79) & (7.37) & (7.13) & (8.01) & (3.91) \\
RP + ANA + ANOM & 0.89 & 1.60 & 0.80 & 0.77 & 0.78 & 0.84 & 1.23 \\
 & (6.97) & (6.62) & (8.35) & (8.17) & (7.32) & (8.66) & (4.05) \\
\bottomrule
\end{tabularx}
\end{center}
\end{table}

\newpage
\begin{table}[!htb]
\caption{Lookahead Bias Analysis: Post-Knowledge Cutoff Portfolios}
\footnotesize
\label{tab:lookahead}
    {\footnotesize This table reports value-weighted decile portfolio performance during periods after the knowledge cutoff of language-model pretraining. Portfolios are formed using each model’s embeddings. Reported are the monthly mean return (Mean), standard deviation (SD), and Sharpe ratio (SR). The analysis spans the following knowledge-cutoff-based windows: $\text{ChronoGPT}_{1999}$ and $\text{ChronoGPT}_{2024}$ are evaluated across the entire out-of-sample, January 2005–June 2024, to enable an apple-for-apple comparison of the same architecture with different knowledge cutoffs. BERT (January 2019–June 2024), RoBERTa (March 2019–June 2024), LLaMA2-13B (October 2022–June 2024), and LLaMA3-8B (April 2023–June 2024). ``H--L'' is a long–short strategy that buys the highest decile (High) and shorts the lowest decile (Low).}
\begin{center}
\begin{tabularx}{\textwidth}{lYYYYYYYYY}
\toprule
& \multicolumn{3}{c}{$\text{ChronoGPT}_{1999}$} & \multicolumn{3}{c}{$\text{ChronoGPT}_{2024}$} & \multicolumn{3}{c}{BERT} \\
\cmidrule(lr){2-4} \cmidrule(lr){5-7} \cmidrule(lr){8-10}
 & Mean & SD & SR & Mean & SD & SR & Mean & SD & SR \\
\midrule
Low (L) & 0.68 & 4.64 & 0.51 & 0.71 & 4.81 & 0.51 & 0.94 & 5.30 & 0.61 \\
2       & 0.66 & 4.21 & 0.54 & 0.76 & 4.67 & 0.57 & 0.95 & 5.53 & 0.60 \\
3       & 0.68 & 4.70 & 0.58 & 0.48 & 4.53 & 0.36 & 1.12 & 5.31 & 0.73 \\
4       & 0.71 & 4.80 & 0.51 & 0.51 & 4.51 & 0.39 & 1.00 & 4.93 & 0.70 \\
5       & 0.82 & 4.81 & 0.67 & 0.72 & 4.65 & 0.58 & 1.30 & 5.02 & 0.89 \\
6       & 0.80 & 4.97 & 0.56 & 0.79 & 4.34 & 0.63 & 1.27 & 5.17 & 0.85 \\
7       & 0.79 & 4.77 & 0.58 & 0.93 & 4.85 & 0.68 & 1.09 & 5.25 & 0.72 \\
8       & 0.86 & 4.93 & 0.61 & 1.01 & 4.76 & 0.73 & 1.15 & 5.47 & 0.73 \\
9       & 0.98 & 5.20 & 0.66 & 1.40 & 6.01 & 1.01 & 1.53 & 5.92 & 0.89 \\
High (H)& 1.64 & 6.19 & 0.92 & 1.68 & 6.68 & 0.84 & 2.70 & 7.41 & 1.26 \\
H--L    & 0.96 & 4.65 & 0.71 & 0.91 & 4.82 & 0.65 & 1.76 & 6.06 & 1.01 \\
\midrule
& \multicolumn{3}{c}{RoBERTa} & \multicolumn{3}{c}{LLaMA2} & \multicolumn{3}{c}{LLaMA3} \\
\cmidrule(lr){2-4} \cmidrule(lr){5-7} \cmidrule(lr){8-10}
 & Mean & SD & SR & Mean & SD & SR & Mean & SD & SR \\
\midrule
Low (L) & 0.63 & 5.16 & 0.42 & 1.67 & 5.03 & 1.15 & 1.13 & 4.06 & 0.96 \\
2       & 0.78 & 5.71 & 0.48 & 0.85 & 3.65 & 0.83 & 0.63 & 3.88 & 0.57 \\
3       & 1.02 & 5.02 & 0.71 & 1.01 & 4.61 & 0.76 & 1.88 & 3.56 & 1.84 \\
4       & 0.78 & 4.97 & 0.54 & 1.25 & 4.52 & 0.96 & 0.93 & 4.94 & 0.65 \\
5       & 0.76 & 5.36 & 0.49 & 1.15 & 4.91 & 0.81 & 2.32 & 5.48 & 1.47 \\
6       & 0.90 & 5.52 & 0.57 & 1.42 & 5.02 & 0.98 & 2.14 & 4.48 & 1.65 \\
7       & 1.25 & 5.35 & 0.81 & 1.66 & 4.74 & 1.22 & 1.00 & 5.53 & 0.63 \\
8       & 1.53 & 5.66 & 0.93 & 1.56 & 5.22 & 1.04 & 0.75 & 5.15 & 0.51 \\
9       & 1.89 & 6.11 & 1.07 & 3.10 & 5.09 & 2.11 & 1.44 & 5.75 & 0.87 \\
High (H)& 2.65 & 7.93 & 1.16 & 3.26 & 7.08 & 1.60 & 3.83 & 7.71 & 1.72 \\
H--L    & 2.01 & 6.67 & 1.05 & 1.59 & 7.39 & 0.74 & 2.70 & 7.30 & 1.28 \\
\bottomrule
\end{tabularx}
\end{center}
\end{table}

\newpage
\begin{landscape}
\begin{table}[!htb]
\footnotesize
\caption{Report-based Forecasts and Characteristics}
\label{tab:chars}
    {\footnotesize This table compares observations in the top (``High $\widehat{\operatorname{Ret}}_{12 m}$'') and bottom (``Low $\widehat{\operatorname{Ret}}_{12 m}$'') deciles of the monthly text-based 12-month return forecast. Panel A reports analyst attributes, and Panel B reports firm characteristics. Means and medians are computed within deciles each month and then averaged over time. $p$-values for differences in means (medians) are from two-sample $t$-tests (Wilcoxon rank-sum tests). Analyst characteristics include forecast accuracy, coverage breadth, experience, broker size, and the Top10 indicator comes from I/B/E/S. Stock returns are from CRSP. Financial variables are from Compustat. See Table \ref{tab:numdef} in the Appendix for detailed variable definitions. The sample spans from January 2005 to December 2023.}
\begin{center}
\begin{tabularx}{1.4\textwidth}{@{\hskip\tabcolsep\extracolsep\fill}l*{6}{c}}
\toprule
\multicolumn{7}{c}{\textbf{Panel A: Analyst Characteristics}}\\
\midrule
 & \multicolumn{2}{c}{High $\widehat{Ret}_{12m}$} & \multicolumn{2}{c}{Low $\widehat{Ret}_{12m}$} & \multicolumn{2}{c}{} \\
\cmidrule{2-3}\cmidrule{4-5}
Variable & Mean & Median & Mean & Median & Difference $p$-Value Means & Difference $p$-Value Medians \\
\midrule
Forecast Accuracy & -0.03 & -0.04 & -0.03 & -0.04 & 0.000 & 0.913 \\
No of industries & 3.45 & 3.00 & 3.50 & 3.00 & 0.005 & 0.000 \\
No of Firms & 19.19 & 18.00 & 21.37 & 19.00 & 0.000 & 0.000 \\
Industry Experience & 16.37 & 16.00 & 16.61 & 16.00 & 0.000 & 0.000 \\
Top10 & 0.32 & 0.00 & 0.33 & 0.00 & 0.000 & 0.000 \\
Broker Size & 68.61 & 48.00 & 73.41 & 55.00 & 0.000 & 0.000 \\
\midrule
\end{tabularx}
\vspace{0.3in}
\begin{tabularx}{1.4\textwidth}{@{\hskip\tabcolsep\extracolsep\fill}l*{6}{c}}
\multicolumn{7}{c}{\textbf{Panel B: Firm Characteristics}}\\
\midrule
 & \multicolumn{2}{c}{High $\widehat{Ret}_{12m}$} & \multicolumn{2}{c}{Low $\widehat{Ret}_{12m}$} & \multicolumn{2}{c}{} \\
\cmidrule{2-3}\cmidrule{4-5}
Variable & Mean & Median & Mean & Median & Difference $p$-Value Means & Difference $p$-Value Medians \\
\midrule
Log Firm Size & 16.18 & 16.15 & 16.06 & 16.05 & 0.000 & 0.000 \\
BM & 0.37 & 0.27 & 0.45 & 0.35 & 0.000 & 0.000 \\
Gross Profit & 0.39 & 0.35 & 0.35 & 0.31 & 0.000 & 0.000 \\
Investment & 1.01 & 0.99 & 0.98 & 0.97 & 0.000 & 0.000 \\
Idiosyncratic Volatility & 0.02 & 0.01 & 0.01 & 0.01 & 0.000 & 0.000 \\
\bottomrule
\end{tabularx}
\end{center}
\end{table}
\end{landscape}

\newpage
\begin{table}[!htb]
\caption{Report-based Forecasts and One-year Abnormal Stock Returns: Characteristics}
\label{tab:char_stats}
\footnotesize
{\footnotesize This table reports mean DGTW characteristic-adjusted buy-and-hold abnormal returns (\%) over 252 trading days for double-sorted portfolios conditioned on analyst- and firm-level characteristics. Detailed characteristics definitions are provided in Table \ref{tab:numdef}. For each characteristic, firms are first sorted into ``Above'' and ``Below'' groups according to the median in the calendar month. Then, within each group, portfolios are formed based on ``High'' and ``Low'' deciles of narrative-based return forecasts. Columns denote Above--High, Above--Low, Below--High, Below--Low, and the difference in spreads across ``Above-High'' minus ``Above-Low'' and ``Below-High'' minus ``Below-Low''. \citet{hodrick1992dividend} t-statistics are reported in parentheses. ***, **, and * indicate significance at the 1\%, 5\%, and 10\% levels, respectively. The sample spans from 2005 to 2023.}
\begin{center}
\begin{tabularx}{\textwidth}{@{\hskip\tabcolsep\extracolsep\fill}lYYYYY}
\toprule
 & \multicolumn{1}{c}{Above-High} & \multicolumn{1}{c}{Above-Low} & \multicolumn{1}{c}{Below-High} & \multicolumn{1}{c}{Below-Low} & \multicolumn{1}{c}{Diff in Spread} \\
\midrule
\multicolumn{6}{l}{\textbf{Panel A: Analyst Characteristics}} \\
\midrule
Forecast Accuracy      & 0.051\sym{***} & -0.013 & 0.043\sym{***} & -0.010 & 0.010 \\
                       & (3.15)         & (-1.43) & (3.98)         & (-1.01) & (0.80) \\
No. of Industries      & 0.062\sym{***} & -0.011 & 0.043\sym{***} & -0.011 & 0.019\sym{*} \\
                       & (3.43)         & (-1.62) & (4.25)         & (-1.06) & (1.80) \\
No. of Firms           & 0.055\sym{***} & -0.005 & 0.043\sym{***} & -0.013 & 0.004 \\
                       & (3.40)         & (-0.46) & (4.09)         & (-1.40) & (0.28) \\
Industry Experience    & 0.066\sym{***} & -0.011 & 0.039\sym{***} & -0.011 & 0.027\sym{*} \\
                       & (4.02)         & (-1.02) & (3.37)         & (-1.16) & (1.82) \\
Top 10                 & 0.034\sym{**}  & -0.012 & 0.050\sym{***} & -0.012 & -0.016 \\
                       & (2.26)         & (-0.96) & (4.37)         & (-1.24) & (-1.33) \\
Broker Size            & 0.056\sym{***} & -0.005 & 0.044\sym{***} & -0.014 & 0.003 \\
                       & (3.13)         & (-0.46) & (4.22)         & (-1.52) & (0.19) \\
\midrule
\multicolumn{6}{l}{\textbf{Panel B: Firm Characteristics}} \\
\midrule
Log Firm Size          & 0.050\sym{***} & -0.012 & 0.026\sym{**}  & -0.006 & 0.030\sym{*} \\
                       & (4.09)         & (-1.43) & (2.17)         & (-0.32) & (1.69) \\
BM                     & 0.001          & -0.020  & 0.061\sym{***} & -0.006 & -0.045\sym{**} \\
                       & (0.07)         & (-1.42) & (3.77)         & (-0.79) & (-2.09) \\
Gross Profit           & 0.067\sym{***} & 0.004   & -0.005         & -0.030\sym{**} & 0.037\sym{*} \\
                       & (3.84)         & (0.45)  & (-0.47)        & (-2.05) & (1.72) \\
Investment             & 0.069\sym{***} & -0.007  & -0.004         & -0.015 & 0.064\sym{***} \\
                       & (3.72)         & (-0.64) & (-0.41)        & (-1.59) & (2.89) \\
Idiosyncratic Volatility & 0.079\sym{***} & 0.014   & 0.016          & -0.017\sym{**} & 0.032 \\
                       & (3.24)         & (0.68)  & (1.48)         & (-2.09) & (1.29) \\
\bottomrule
\end{tabularx}
\end{center}
\end{table}

\newpage
\begin{table}[!htb]
\caption{Factor Loadings}
\label{tab:alpha_ff}
\footnotesize
    {\footnotesize This table reports factor regressions of value-weighted long–short (L–S) and long-only portfolios sorting on report-based return forecasts. Alphas and factor loadings are from regressions on the \citet{fama2018choosing} six factors. t-statistics (in parentheses) are Newey–West adjusted with 12 lags. ***, **, and * indicate significance at the 1\%, 5\%, and 10\% levels, respectively. The sample spans from January 2005 to June 2024.}
\begin{center}
\begin{tabularx}{\textwidth}{@{\hskip\tabcolsep\extracolsep\fill}lYYYYYY}
\toprule
  & \multicolumn{3}{c}{L-S}      & \multicolumn{3}{c}{Long only}   \\
\cmidrule(lr){2-4}\cmidrule(lr){5-7}
            &\multicolumn{1}{c}{(1)}&\multicolumn{1}{c}{(2)}&\multicolumn{1}{c}
            {(3)}&\multicolumn{1}{c}{(4)}&\multicolumn{1}{c}{(5)}&\multicolumn{1}{c}{(6)}\\
\midrule
Alpha  & 1.026\sym{**} & 0.644\sym{**} & 0.666\sym{**}
       & 1.543\sym{***} & 0.504\sym{***} & 0.527\sym{**} \\
       & (2.57) & (2.50) & (2.54) & (3.62) & (2.63) & (2.55) \\
MKT-RF &         & 0.369\sym{***} & 0.270\sym{***}
       &         & 1.243\sym{***} & 1.195\sym{***} \\
       &         & (5.37) & (3.88) &         & (25.12) & (22.94) \\
SMB    &         & 0.098 & 0.117
       &         & 0.001 & 0.002 \\
       &         & (0.68) & (0.77) &         & (0.01) & (0.02) \\
HML    &         & -0.940\sym{***} & -0.707\sym{***}
       &         & -0.483\sym{***} & -0.346\sym{***} \\
       &         & (-7.19) & (-4.61) &         & (-4.90) & (-3.73) \\
RMW    &         &         & 0.311
       &         &         & 0.119 \\
       &         &         & (1.62) &         &         & (1.19) \\
CMA    &         &         & -0.789\sym{***}
       &         &         & -0.423\sym{**} \\
       &         &         & (-3.11) &         &         & (-2.35) \\
MOM    &         &         & -0.136
       &         &         & -0.049 \\
       &         &         & (-1.62) &         &         & (-0.56) \\
\midrule
Months & 234 & 234 & 234 & 234 & 234 & 234 \\
Adj. $R^2$ & 0.000 & 0.369 & 0.438 & 0.000 & 0.782 & 0.792 \\
\bottomrule
\end{tabularx}
\end{center}
\end{table}

\newpage
\begin{table}[!htb]
\caption{Contemporaneous News and Characteristics of Decile Portfolios}
\footnotesize
\label{tab:sentiment}
    {\footnotesize This table presents deciles of analyst reports sorted by ridge models' return forecasts. For each decile, I report the predicted and realized 12-month ahead returns, and five measures. $CAR_{[0,+1]}$ is two-day DGTW characteristic-adjusted buy-and-hold abnormal returns following report release dates (in \%). $SUE$ is the latest earnings surprise prior to the report release, calculated as the actual EPS minus the last consensus EPS forecast before the earnings announcement. $REC_{REV}$ denotes recommendation revision, calculated as the current report's recommendation minus the last recommendation in I/B/E/S issued by the same analyst for the same stock. $EF_{REV}$ refers to earnings forecast revision, calculated as the current report's EPS forecast minus the last EPS forecast in I/B/E/S issued by the same analyst for the same stock, scaled by the stock price 50 days before the report release. $TP_{REV}$ represents the target price revision, calculated as the current report's target price minus the last target price in I/B/E/S issued by the same analyst for the same stock, scaled by the stock price 50 days before the report release. *** denotes statistical significance at the 1\% level.}
    \begin{center}
\begin{tabularx}{1.0\textwidth}{lYYYYYYY}
\toprule
Decile & $\widehat{Ret}_{12m}$ & $Ret_{12m}$ & $CAR_{[0,+1]}$ & SUE & $REC_{REV}$ & $EF_{REV}$ & $TP_{REV}$ \\
\midrule
1 & -0.178 & 0.115 & 0.163 & 0.001 & 0.007 & 0.001 & 0.012 \\
2 & -0.042 & 0.109 & 0.102 & 0.001 & 0.004 & 0.001 & 0.013 \\
3 & 0.014 & 0.110 & 0.133 & -0.001 & 0.001 & 0.000 & 0.011 \\
4 & 0.057 & 0.112 & 0.108 & -0.001 & -0.001 & 0.001 & 0.009 \\
5 & 0.095 & 0.116 & 0.089 & -0.001 & -0.002 & 0.000 & 0.008 \\
6 & 0.132 & 0.119 & 0.065 & -0.001 & 0.001 & -0.000 & 0.007 \\
7 & 0.170 & 0.124 & 0.046 & -0.001 & -0.002 & 0.000 & 0.006 \\
8 & 0.214 & 0.133 & 0.031 & -0.003 & -0.004 & -0.000 & 0.004 \\
9 & 0.273 & 0.139 & 0.004 & -0.004 & -0.003 & -0.001 & 0.003 \\
10 & 0.418 & 0.164 & -0.087 & -0.006 & -0.003 & -0.002 & -0.001 \\
H-L & 0.596\sym{***} & 0.049\sym{***} & -0.250\sym{***} & -0.007\sym{***} & -0.010\sym{***} & -0.003\sym{***} & -0.013\sym{***} \\
\bottomrule
\end{tabularx}
    \end{center}
\end{table}

\newpage
\begin{table}[!htb]
\caption{Sources of Investment Value}
\label{tab:shap}
\footnotesize
    {\footnotesize This table presents an analysis of analyst report topics across five categories, examining both their distribution and investment value contribution. The categories cover Financial Analysis, Company and Industry Overview, Strategic Outlook, Risk and Governance, and Additional Content. Panel A displays aggregate statistics on topic distribution, showing the total number of sentences (in millions) and tokens (in billions) per category, along with their respective percentages of the total content. Panel B quantifies each category's contribution to portfolio performance through Shapley value decomposition, specifically examining both the Sharpe ratios and returns of the value-weighted ``H-L'' portfolio.}
    \begin{center}
    \begin{tabularx}{\textwidth}{p{6cm}YYYY}
    \toprule
    \multicolumn{5}{l}{Panel A: Topic Distribution} \\
    \midrule
    Category & \#Sentences & \#Tokens & \%Sentences & \%Tokens \\
    \midrule
    Financial Analysis & 19.14 & 0.61 & 36.56 & 36.57 \\
    Company and Industry Overview & 14.94 & 0.48 & 28.53 & 28.75 \\
    Strategic Outlook & 7.92 & 0.26 & 15.13 & 15.61 \\
    Risk and Governance & 7.40 & 0.24 & 14.14 & 14.35 \\
    Additional Content & 2.95 & 0.08 & 5.63 & 4.71 \\
    \bottomrule
    \end{tabularx}
    \begin{tabularx}{\textwidth}{p{6cm}YYYY}
    \multicolumn{5}{l}{Panel B: Shapley Value Decomposition} \\
    \midrule
    Category & SHAP(SR) & SHAP(Ret) & \%SHAP(SR) & \%SHAP(Ret) \\
    \midrule
    Strategic Outlook & 0.24 & 0.28 & 41.34 & 31.43 \\
    Company and Industry Overview & 0.16 & 0.24 & 27.61 & 26.92 \\
    Risk and Governance & 0.06 & 0.19 & 11.21 & 21.36 \\
    Financial Analysis & 0.09 & 0.18 & 16.39 & 19.53 \\
    Additional Content & 0.02 & 0.01 & 3.44 & 0.77 \\
    \bottomrule
    \end{tabularx}
    \end{center}
\end{table}

\newpage
\begin{table}[!htb]
\caption{Sources of Investment Value in Strategic Outlook}
\footnotesize
\label{tab:soshap}
    {\footnotesize This table analyzes analysts' strategic outlook discussions across three dimensions, quantifying their contribution to investment value through Shapley analysis. Panel A examines the temporal dimension, decomposing sentences into long-term, short-term, and combined timeframe categories. Panel B classifies sentences by sentiment orientation (negative, neutral, or positive), while Panel C categorizes content based on discussion focus (risk versus fundamental analysis). For each dimension and category, I report both absolute Shapley values (SHAP(SR) and SHAP(Ret)) and their relative percentages, measuring their contributions to portfolio Sharpe ratio and returns, respectively.}
    \begin{center}
    \begin{tabularx}{\textwidth}{p{3.5cm}YYYY}
    \toprule
    Dimension & SHAP(SR) & SHAP(Ret) & \%SHAP(SR) & \%SHAP(Ret) \\
    \midrule
    \multicolumn{5}{l}{Panel A: Timeframe} \\
    \midrule
    Long-term & 0.39 & 0.52 & 49.51 & 50.20 \\
    Short-term & 0.21 & 0.28 & 26.69 & 26.70 \\
    Both & 0.19 & 0.24 & 23.80 & 23.10 \\
    \midrule
    \multicolumn{5}{l}{Panel B: Sentiment} \\
    \midrule
    Negative & 0.19 & 0.21 & 24.34 & 20.23 \\
    Neutral & 0.20 & 0.30 & 25.72 & 28.91 \\
    Positive & 0.39 & 0.53 & 49.94 & 50.86 \\
    \midrule
    \multicolumn{5}{l}{Panel C: Focus} \\
    \midrule
    Risk & 0.10 & 0.13 & 13.00 & 12.25 \\
    Fundamental & 0.68 & 0.92 & 87.00 & 87.75 \\
    \bottomrule
    \end{tabularx}
    \end{center}
\end{table}

\newpage
\begin{table}[!htb]
\footnotesize
\caption{Category Portfolio Statistics}
\label{tab:catdecile}
    {\footnotesize This table reports the value-weighted decile portfolio performances sorted by ridge models' forecast using the past 12 months' analyst report category-specific content. Categories include strategic outlook, company and industry overview, financial analysis, and risk and governance. The prediction target is the next 12 months' return of the stock. For each portfolio, I report several performance measures: excess return mean and standard deviation, Sharpe ratio, $\alpha$ relative to \citet{fama2018choosing} six factors, and the corresponding t-statistics for $\alpha$. Portfolios are rebalanced monthly using the average of the most recent 12 months of out-of-sample predictions. The `H-L' row represents a long-short strategy that takes a long position in the highest decile (High) and a short position in the lowest decile (Low). The sample period extends from January 2005 to June 2024.}
    \begin{center}
    \scalebox{1}{
    \begin{tabularx}{\textwidth}{@{\hskip\tabcolsep\extracolsep\fill}l*{10}r}
    \toprule
    & \multicolumn{5}{c}{Strategic Outlook} & \multicolumn{5}{c}{Company and Industry Overview} \\
    \cmidrule{2-6} \cmidrule{7-11}
     & Mean & SD & SR & $\alpha$ & $t_\alpha$  & Mean & SD & SR & $\alpha$ & $t_\alpha$ \\
    \midrule
    Low (L) & 0.46 & 4.32 & 0.37 & -0.29 & -2.18 & 0.76 & 5.69 & 0.46 & 0.08 & 0.38 \\
    2 & 0.81 & 4.26 & 0.66 & 0.10 & 1.03 & 0.56 & 4.57 & 0.43 & -0.14 & -1.15 \\
    3 & 0.73 & 4.85 & 0.52 & -0.00 & -0.03 & 0.61 & 4.28 & 0.49 & -0.09 & -1.25 \\
    4 & 0.71 & 4.62 & 0.54 & -0.10 & -1.04 & 0.59 & 4.37 & 0.47 & -0.15 & -2.08 \\
    5 & 0.58 & 4.94 & 0.40 & -0.19 & -1.43 & 0.75 & 4.54 & 0.57 & -0.02 & -0.20 \\
    6 & 0.68 & 4.84 & 0.49 & -0.06 & -0.69 & 0.79 & 4.83 & 0.56 & 0.01 & 0.14 \\
    7 & 0.79 & 4.92 & 0.55 & -0.00 & -0.03 & 0.97 & 4.67 & 0.72 & 0.17 & 1.50 \\
    8 & 0.87 & 5.07 & 0.59 & 0.08 & 0.63 & 0.81 & 4.74 & 0.59 & -0.06 & -0.46 \\
    9 & 1.08 & 4.92 & 0.76 & 0.25 & 1.74 & 0.94 & 4.79 & 0.68 & 0.08 & 0.85 \\
    High (H) & 1.87 & 6.65 & 0.97 & 0.72 & 2.85 & 1.33 & 5.97 & 0.77 & 0.29 & 1.71 \\
    H-L & 1.41 & 5.25 & 0.93 & 1.00 & 3.18 & 0.57 & 5.23 & 0.38 & 0.21 & 0.92 \\
    \end{tabularx}}
    \vspace{0.3in}   
    \scalebox{1
    }{
    \begin{tabularx}{\textwidth}{@{\hskip\tabcolsep\extracolsep\fill}l*{10}r}
    \toprule
    & \multicolumn{5}{c}{Financial Analysis} & \multicolumn{5}{c}{Risk and Governance} \\
    \cmidrule{2-6} \cmidrule{7-11}
     & Mean & SD & SR & $\alpha$ & $t_\alpha$  & Mean & SD & SR & $\alpha$ & $t_\alpha$ \\
    \midrule
    Low (L) & 0.82 & 5.03 & 0.56 & 0.14 & 0.66 & 0.85 & 4.97 & 0.59 & 0.20 & 1.01 \\
    2 & 0.53 & 4.26 & 0.43 & -0.17 & -1.71 & 0.63 & 4.50 & 0.48 & -0.01 & -0.10 \\
    3 & 0.52 & 4.60 & 0.39 & -0.24 & -2.38 & 0.49 & 4.49 & 0.38 & -0.23 & -2.63 \\
    4 & 0.73 & 4.65 & 0.54 & 0.00 & 0.03 & 0.77 & 4.69 & 0.57 & 0.02 & 0.23 \\
    5 & 0.63 & 4.52 & 0.48 & -0.12 & -1.49 & 0.70 & 4.65 & 0.52 & -0.05 & -0.59 \\
    6 & 0.93 & 4.62 & 0.70 & 0.19 & 1.59 & 0.82 & 4.55 & 0.63 & 0.04 & 0.41 \\
    7 & 0.81 & 4.80 & 0.58 & -0.03 & -0.28 & 0.72 & 4.77 & 0.52 & -0.11 & -1.34 \\
    8 & 0.75 & 4.82 & 0.54 & -0.08 & -0.80 & 0.94 & 4.82 & 0.68 & 0.09 & 0.84 \\
    9 & 1.07 & 4.99 & 0.74 & 0.17 & 1.33 & 1.06 & 4.89 & 0.75 & 0.15 & 1.34 \\
    High (H) & 1.39 & 6.09 & 0.79 & 0.34 & 1.71 & 1.37 & 5.90 & 0.81 & 0.32 & 1.97 \\
    H-L & 0.58 & 4.81 & 0.41 & 0.19 & 0.74 & 0.52 & 4.90 & 0.37 & 0.12 & 0.44 \\
    \bottomrule
    \end{tabularx}}
    \end{center}
\end{table}




\appendix 

\setcounter{page}{1}
\setcounter{table}{0}
\setcounter{figure}{0}

\renewcommand{\thetable}{A\arabic{table}}
\renewcommand{\thefigure}{A\arabic{figure}}

\newpage
\section*{A. Additional Tables and Figures}
\captionsetup[subfigure]{labelformat=empty, justification=centering}

\begin{figure}[H]
\begin{center}
\begin{subfigure}[b]{0.48\textwidth}
\includegraphics[width=\textwidth]{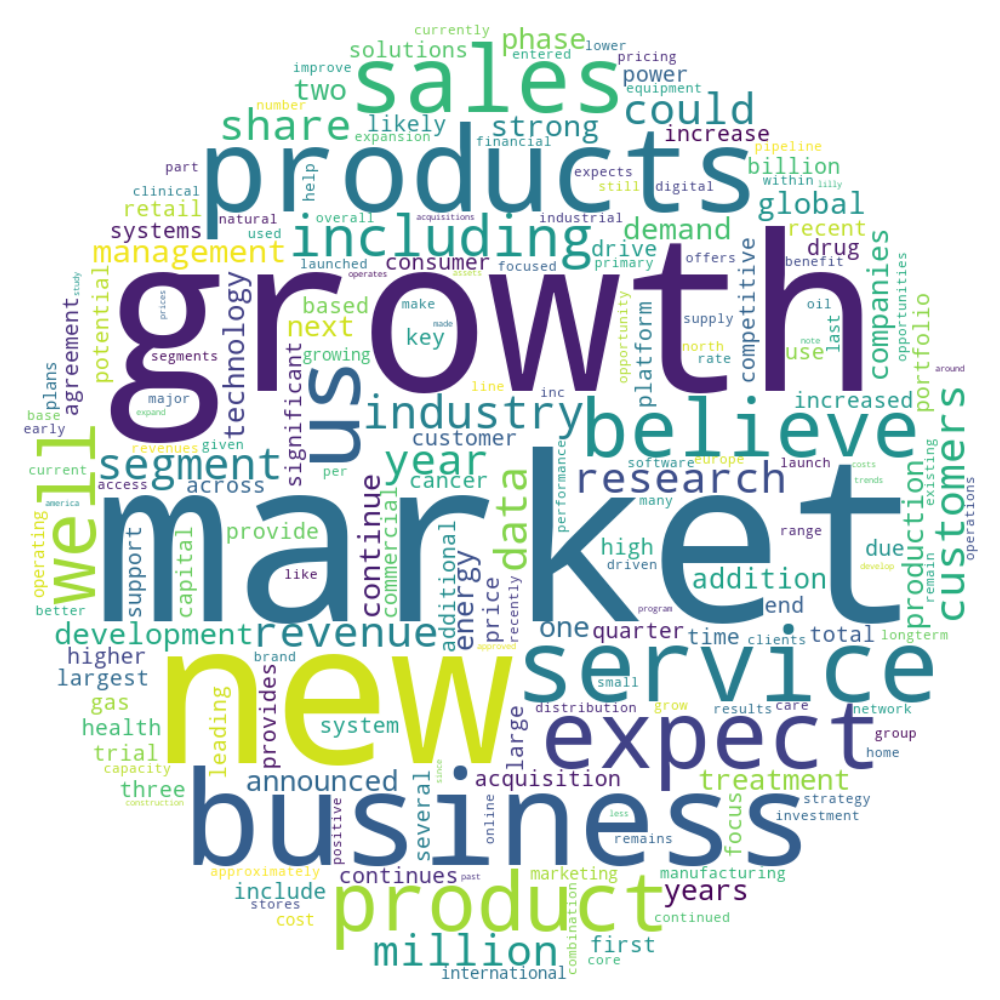}
\caption{(1) Company and Industry Overview}
\end{subfigure}
\hfill
\begin{subfigure}[b]{0.48\textwidth}
\includegraphics[width=\textwidth]{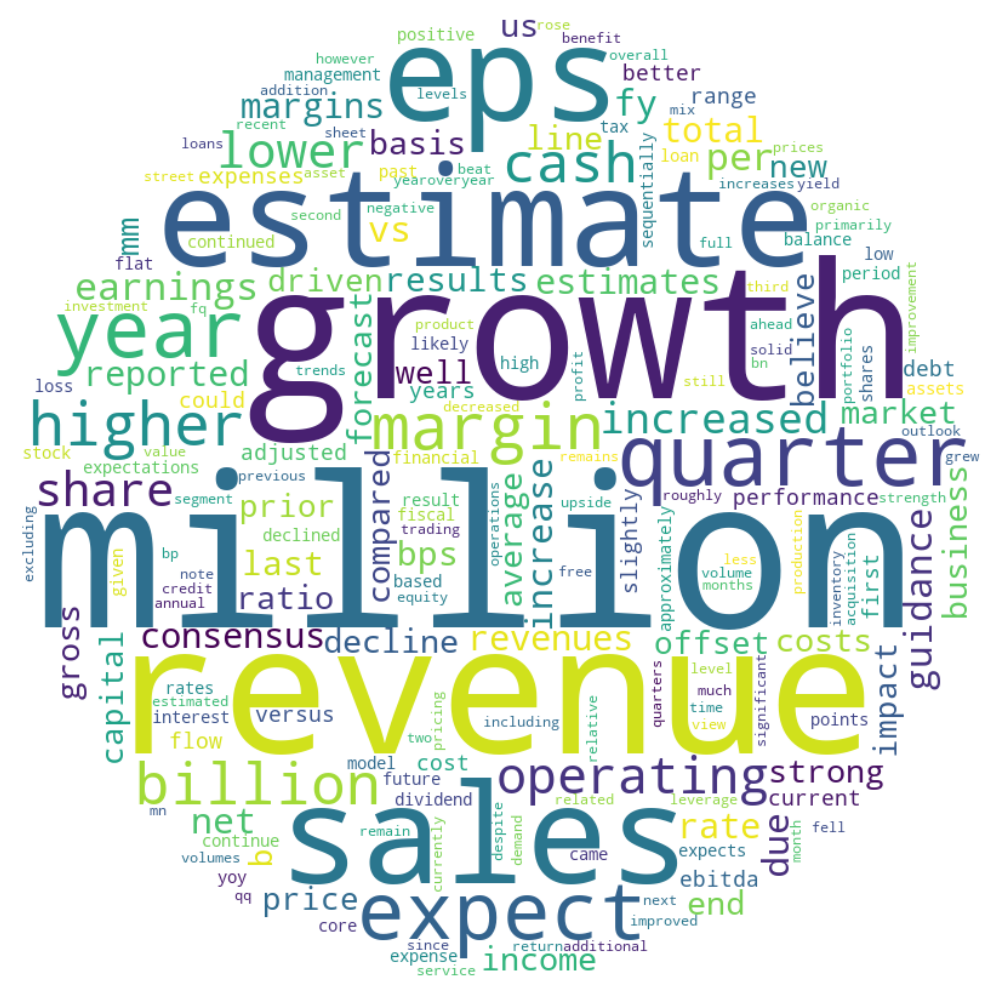}
\caption{(2) Financial Analysis}
\end{subfigure}
\vfill
\vspace{0.1in}
\begin{subfigure}[b]{0.48\textwidth}
\includegraphics[width=\textwidth]{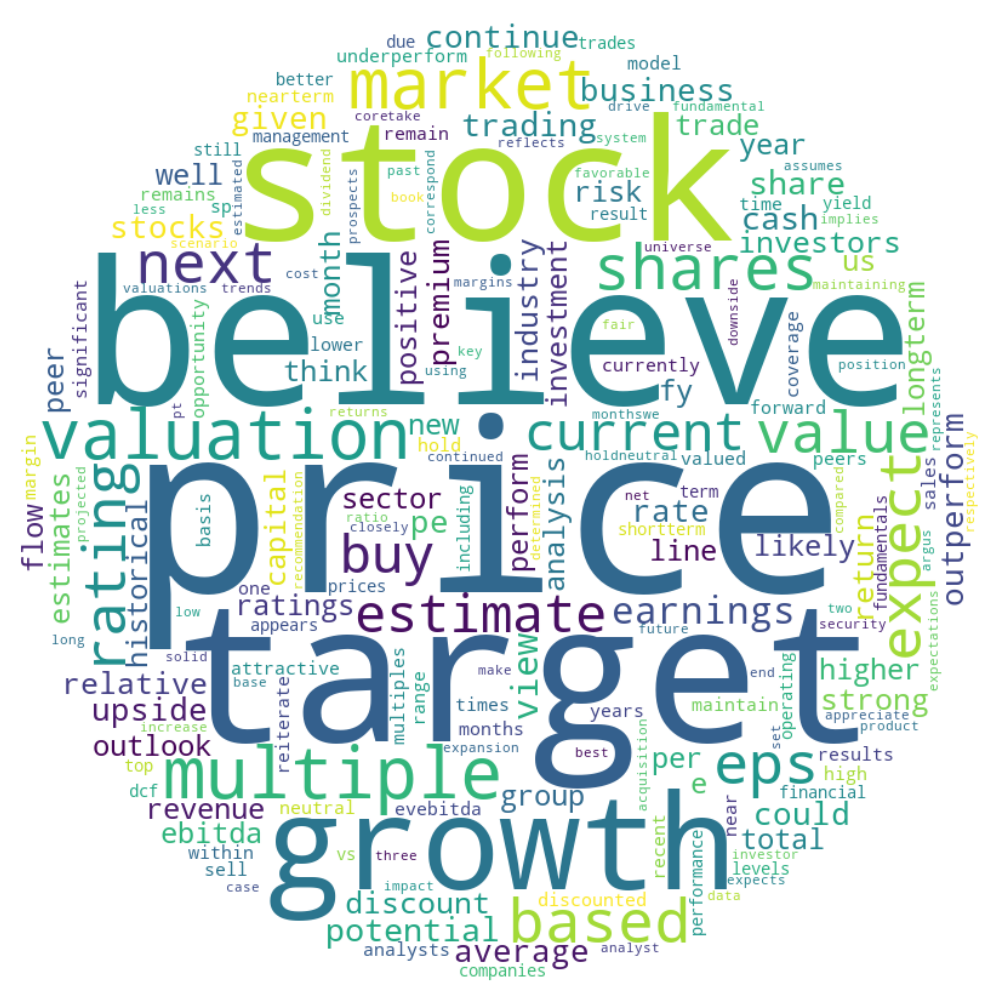}
\caption{(3) Strategic Outlook}
\end{subfigure}
\hfill
\begin{subfigure}[b]{0.48\textwidth}
\includegraphics[width=\textwidth]{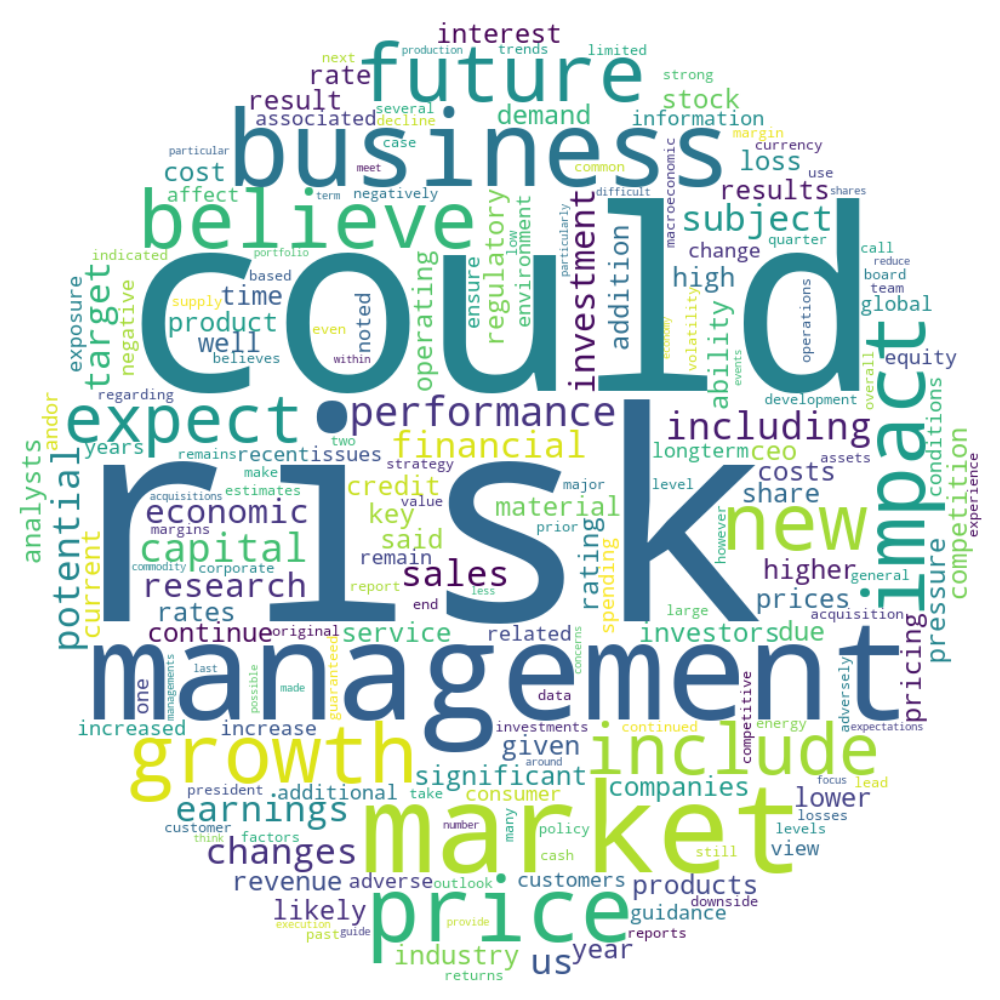}
\caption{(4) Risk and Governance}
\end{subfigure}
\end{center}
\caption{Word Clouds of Topics}\label{fig: wordcloud}
\vspace{0.2in}
    {\footnotesize Note: This figure presents word clouds for 4 categories frequently discussed in analyst reports. Each word cloud visually represents the most common terms associated with the topic, with word size indicating term frequency. The categories include Company and Industry Overview, Financial Analysis, Strategic Outlook, and Risk and Governance. The Additional Content category is excluded from the visualization.}
\end{figure}

\newpage
\begin{figure}[!htb]
\captionsetup{skip=-0.5em}
\caption{Report-based Forecasts and Abnormal Stock Returns: Alternative LLMs} \label{fig:car_alterllm}
\begin{center}
	\includegraphics[width = \textwidth]{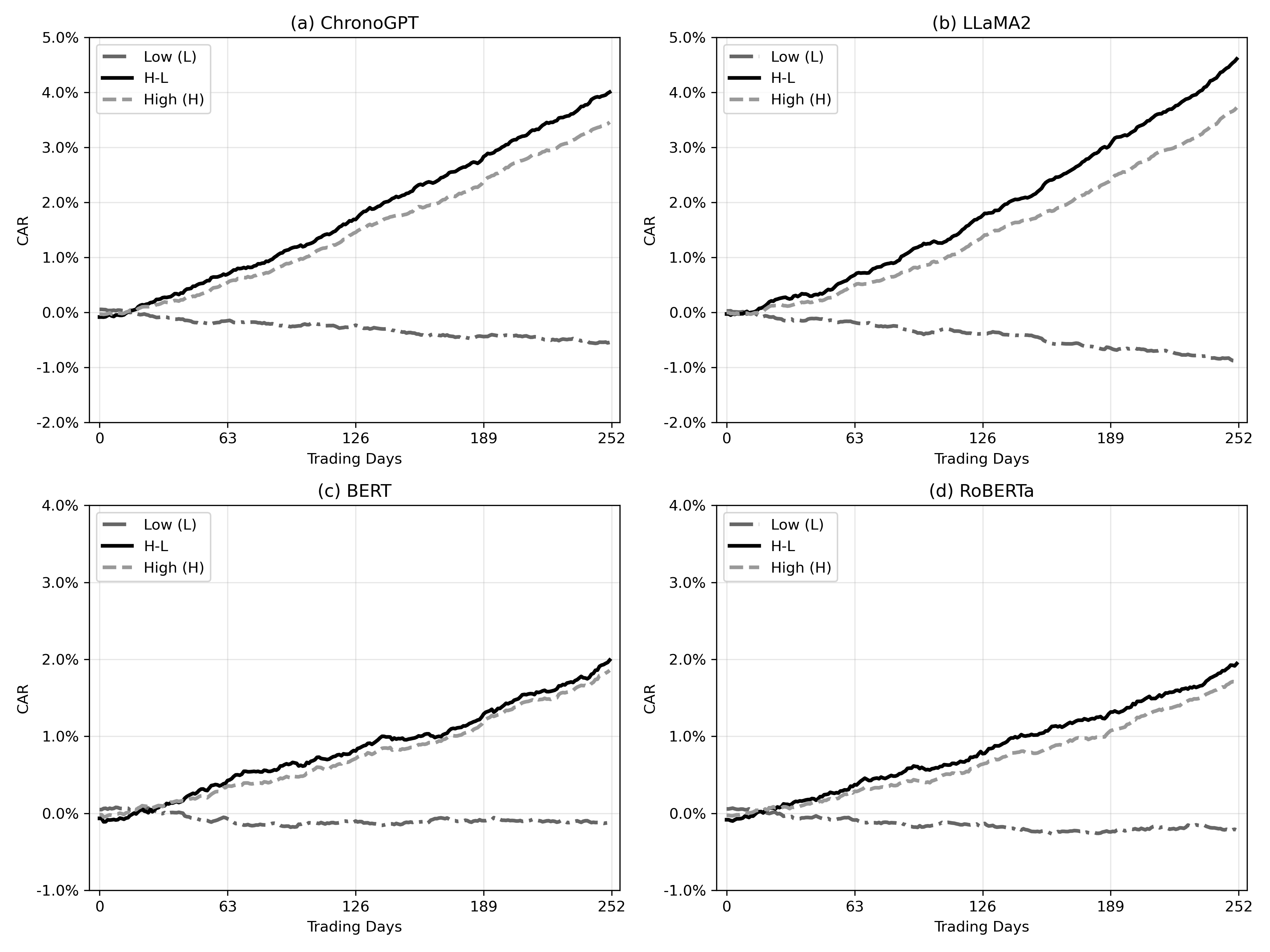}
\end{center}
\vspace{-0.2in}
{\footnotesize Note: This figure presents DGTW characteristic-adjusted buy-and-hold abnormal returns following research report releases, measured over a [0, +252] trading day window after each announcement. Reports are sorted into deciles based on the predicted 12-month returns using four LLMs: (a) $\text{ChronoGPT}_{1999}$, (b) LLaMa2-13B, (c) BERT, (d) RoBERTa. The `High (H)' group consists of reports predicting returns in the highest decile within each trading day, while the `Low (L)' group comprises reports predicting returns in the lowest decile. Returns are value-weighted within each calendar month using lagged market-capitalization weights and then equally averaged across months. `H-L' denotes the return spread between the High and Low groups. The sample spans from 2005 to 2023.}
\end{figure}

\newpage
\begin{figure}[!htb]
\captionsetup{skip=-0.5em}
\caption{Report-based Forecasts and Abnormal Stock Returns: Analyst Characteristics} \label{fig:xs_ana}
\begin{center}
	\includegraphics[width =\textwidth]{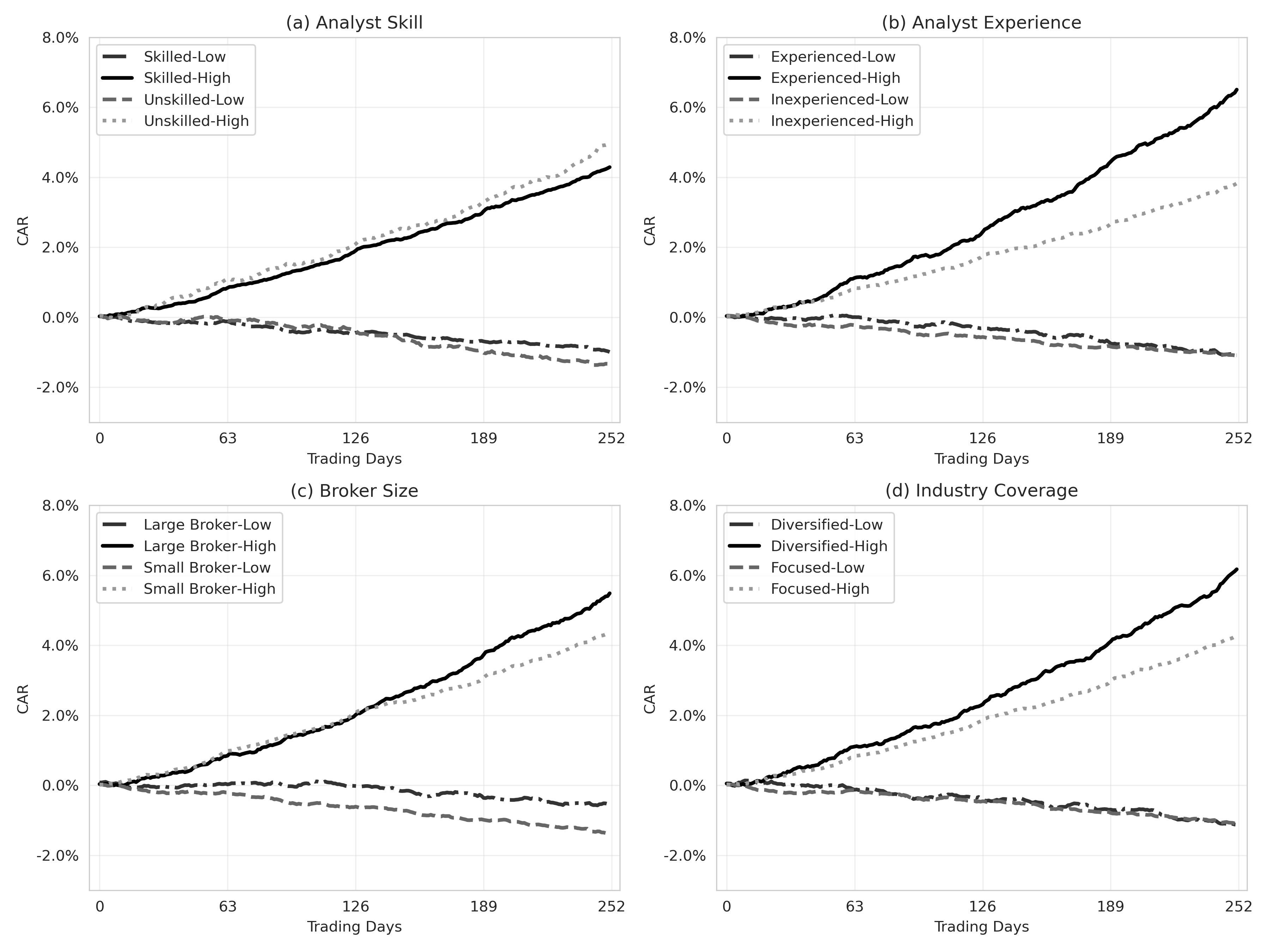}
\end{center}
\vspace{-0.2in}
{\footnotesize Note: This figure presents DGTW characteristic-adjusted buy-and-hold abnormal returns following research report releases, measured over a [0, +252] trading day window after each announcement. Reports are sorted based on two dimensions. First, they are categorized by the monthly median of the following characteristics at the end of the previous month: (a) analyst forecast accuracy, (b) analyst industry experience, (c) brokerage firm size, and (d) analyst industry coverage breadth. Second, within each category, reports are sorted into deciles based on their predicted 12-month returns. Returns are value-weighted within each calendar month using lagged market-capitalization weights and then equally averaged across months. The sample spans from 2005 to 2023.}
\end{figure}

\newpage

\newpage
\begin{table}[!htb]
\caption{Variables Description}\label{tab:numdef}
{\footnotesize This table shows the definitions of numerical measures.}
\begin{center}
\scriptsize
\tabcolsep = 0.45cm
\renewcommand{\arraystretch}{1.2}
\begin{tabularx}{\textwidth}{@{}p{4.5cm}p{11cm}@{}}
\toprule
Numerical Measures & Definition and/or sources \\
\midrule
\textbf{Panel A: Firm-Level} & \\
Log Firm Size &  The logarithm of market value equity of the firm ($CSHOQ * PRCCQ$) at the end of the month prior to report release. \\
BM & The book value of equity ($SEQ + TXDB + ITCB -PREF$) scaled by the market value of equity ($CSHOQ * PRCCQ$) t the end of the month prior to the report release. \\
SUE & Earnings surprise, calculated as the actual EPS minus the last consensus EPS forecast before the earnings announcement. Consensus EPS is the median value of 1-year EPS forecasts within a 90-day window of all analysts following the firm. The unexpected earnings is scaled by price per share at the fiscal quarter end.\\
Gross Profit &  Revenue (sale) - cost of goods solds (cogs), divided by total assets (at).  \\
Investment &   Ratio of capital investment (capx) to revenue (revt) divided by the firm-specific 36-month rolling mean of that ratio. Exclude if revenue less than \$10m. \\
Idiosyncratic Volatility &  Standard deviation of residuals from Fama-French three-factor regressions using the past month of daily data. \\
\midrule
\textbf{Panel B: Analyst-Level} & \\
Forecast Accuracy & The proportional mean absolute forecast error calculated as the difference between the absolute forecast error (AFE) for analyst $i$ on firm $j$ at time $t$ and the mean absolute forecast error ($MAFE$) for firm $j$ at time $t$ scaled by the mean absolute forecast error for firm $j$ at time $t$. \\
No of industries & The number of Fama–French 48 industries covered by the analyst at year $t$. \\
No of Firms & The number of firms followed by the analyst at year $t$. \\
Industry Experience &  Analyst‑year measure equal to the number of calendar years since the analyst’s first appearance in I/B/E/S. \\
Top10 &  Indicator equal to 1 if a brokerage ranks in the top ten by Broker Size within year $t$, and 0 otherwise. (Ranking is computed cross-sectionally each year.) \\
Broker Size & The number of distinct analysts employed by a brokerage in a calendar year, proxied by the count of unique analyst–broker pairs with at least one EPS forecast in I/B/E/S that year \\
\midrule
\textbf{Panel C: Report-Level} & \\
$REC_{REV}$ & Recommendation revision, calculated as the current report's recommendation minus the last recommendation in I/B/E/S issued by the same analyst for the same firm. \\
$EF_{REV}$ & Earnings forecast revision, calculated as the current report's EPS forecast minus the last EPS forecast in I/B/E/S issued by the same analyst for the same firm, scaled by the stock price 50 days before the report release. \\
$TP_{REV}$ & Target price revision, calculated as the current report's target price minus the last target price in I/B/E/S issued by the same analyst for the same firm, scaled by the stock price 50 days before the report release. \\
Tone & Net sentiment of the report text, computed by classifying each sentence with a fine-tuned BERT model as positive, negative, or neutral; defined at the report level as Tone ${ }_{j t}=\left(N_{j t}^{+}-N_{j t}^{-}\right) / N_{j t}$, where $N_{j t}^{+}$and $N_{j t}^{-}$are counts of positive and negative sentences and $N_{j t}$ is the total number of sentences. \\
\bottomrule
\end{tabularx}
\end{center}
\end{table}

\newpage
\begin{table}[!htb]
\caption{Characteristics-based Predictors}\label{anom}
{\footnotesize This table reports the market and firm-related
fundamental factors considered in this paper. I follow \citet{gu2020empirical} to construct the dataset. \citet{gu2020empirical} provide a detailed description of the dataset in their appendix.}
\begin{center}
\scriptsize
\tabcolsep = 0.22cm
\renewcommand{\arraystretch}{1.2}
\begin{tabularx}{\textwidth}{llll}
\toprule
 Acronym &                                Description &         Acronym &                                        Description \\
\midrule
  absacc &                          Absolute accruals &          mom36m &                                  36-month momentum \\
     acc &                   Working capital accruals &           mom6m &                                   6-month momentum \\
  aeavol &      Abnormal earnings announcement volume &              ms &                          Financial statement score \\
     age &     \# years since first Compustat coverage &           mvel1 &                                               Size \\
     agr &                               Asset growth &          mve\_ia &                             Industry-adjusted size \\
baspread &                             Bid-ask spread &          niincr &                       Number of earnings increases \\
    beta &                                       Beta &        operprof &                            Operating profitability \\
  betasq &                               Beta squared &          orgcap &                             Organizational capital \\
      bm &                             Book-to-market &      pchcapx\_ia & Industry adjusted \% change in capital expenditures \\
   bm\_ia &           Industry-adjusted book to market &       pchcurrat &                          \% change in current ratio \\
    cash &                              Cash holdings &         pchdepr &                           \% change in depreciation \\
cashdebt &                          Cash flow to debt &   pchgm\_pchsale &       \% change in gross margin - \% change in sales \\
  cashpr &                          Cash productivity &        pchquick &                            \% change in quick ratio \\
     cfp &                   Cash flow to price ratio & pchsale\_pchinvt &          \% change in sales - \% change in inventory \\
  cfp\_ia & Industry-adjusted cash flow to price ratio & pchsale\_pchrect &                \% change in sales - \% change in A/R \\
 chatoia & Industry-adjusted change in asset turnover & pchsale\_pchxsga &               \% change in sales - \% change in SG\&A \\
  chcsho &               Change in shares outstanding &      pchsaleinv &                     \% change in sales-to-inventory \\
 chempia &      Industry-adjusted change in employees &          pctacc &                                   Percent accruals \\
   chinv &                        Change in inventory &      pricedelay &                                        Price delay \\
   chmom &                 Change in 6-month momentum &              ps &                         Financial statements score \\
  chpmia &  Industry-adjusted change in profit margin &           quick &                                        Quick ratio \\
    chtx &                      Change in tax expense &              rd &                                       R\&D increase \\
 cinvest &                       Corporate investment &          rd\_mve &                       R\&D to market capitalization \\
 convind &                 Convertible debt indicator &         rd\_sale &                                       R\&D to sales \\
  currat &                              Current ratio &      realestate &                               Real estate holdings \\
    depr &                        Depreciation / PP\&E &          retvol &                                  Return volatility \\
    divi &                        Dividend initiation &            roaq &                                   Return on assets \\
    divo &                          Dividend omission &          roavol &                                Earnings volatility \\
  dolvol &                      Dollar trading volume &            roeq &                                   Return on equity \\
      dy &                          Dividend to price &            roic &                         Return on invested capital \\
     ear &               Earnings announcement return &            rsup &                                   Revenue surprise \\
     egr &        Growth in common shareholder equity &        salecash &                                      Sales to cash \\
      ep &                          Earnings to price &         saleinv &                                 Sales to inventory \\
     gma &                        Gross profitability &         salerec &                               Sales to receivables \\
  grCAPX &             Growth in capital expenditures &         secured &                                       Secured debt \\
 grltnoa &   Growth in long term net operating assets &      securedind &                             Secured debt indicator \\
    herf &               Industry sales concentration &             sgr &                                       Sales growth \\
    hire &                       Employee growth rate &             sin &                                         Sin stocks \\
 idiovol &            Idiosyncratic return volatility &              sp &                                     Sales to price \\
     ill &                                Illiquidity &      std\_dolvol &    Volatility of liquidity (dollar trading volume) \\
  indmom &                          Industry momentum &        std\_turn &           Volatility of liquidity (share turnover) \\
  invest &         Capital expenditures and inventory &          stdacc &                                 Accrual volatility \\
     lev &                                   Leverage &           stdcf &                               Cash flow volatility \\
     lgr &                   Growth in long-term debt &            tang &                     Debt capacity/firm tangibility \\
  maxret &                       Maximum daily return &              tb &                          Tax income to book income \\
  mom12m &                          12-month momentum &            turn &                                     Share turnover \\
   mom1m &                           1-month momentum &       zerotrade &                                  Zero trading days \\
\bottomrule
\end{tabularx}
\end{center}
\end{table}

\newpage
\begin{table}[!htb]
\caption{Analyst-based Predictors}\label{ananum}
{\footnotesize This table reports the analyst factors considered in this paper. I compile the data from \cite{ChenZimmermann2021} website.}
\begin{center}
\scriptsize
\begin{tabular}{p{0.3\textwidth}p{0.2\textwidth}p{0.4\textwidth}}
\toprule
Acronym & Journal (Publish Year) & Description \\
\midrule
AnalystRevision & FAJ (1984) & EPS forecast revision \\
AnalystValue & JAE (1998) & Analyst Value \\
AOP & JAE (1998) & Analyst Optimism \\
ChangeInRecommendation & JF (2004) & Change in recommendation \\
ChForecastAccrual & RAS (2004) & Change in Forecast and Accrual \\
ChNAnalyst & ROF (2008) & Decline in Analyst Coverage \\
ConsRecomm & JF (2001) & Consensus Recommendation \\
CredRatDG & JF (2001) & Credit Rating Downgrade \\
DownRecomm & JF (2001) & Down forecast EPS \\
EarningsForecastDisparity & JFE (2011) & Long-vs-short EPS forecasts \\
ExclExp & RAS (2003) & Excluded Expenses \\
FEPS & WP (2006) & Analyst earnings per share \\
fgr5yrLag & JF (1996) & Long-term EPS forecast \\
ForecastDispersion & JF (2002) & EPS Forecast Dispersion \\
Recomm\_ShortInterest & AR (2011) & Analyst Recommendations and Short-Interest \\
REV6 & JF (1996) & Earnings forecast revisions \\
sfe & AR (2001) & Earnings Forecast to price \\
UpRecomm & JF (2001) & Up Forecast \\
\bottomrule
\end{tabular}
\end{center}
\end{table}

\newpage
\scriptsize
\renewcommand{\arraystretch}{1.2}
\begin{longtable}{>{\raggedright\arraybackslash}p{4cm}>
{\raggedright\arraybackslash}p{12cm}}
\caption{Topic Categories and Descriptions}\label{tab: topic} \\
\toprule
\textbf{Topic} & \textbf{Desriptions} \\
\midrule
\endfirsthead
\multicolumn{2}{c}{{\tablename\ \thetable{} -- continued from previous page}} \\
\toprule
\textbf{Topic} & \textbf{Desriptions} \\
\midrule
\endhead
\endfoot
\bottomrule
\endlastfoot
\multicolumn{2}{l}{\textbf{Company and Industry Overview}} \\
Executive Summary & Provides a high-level overview of the report's key findings and conclusions; includes a brief description of the company, its industry, and the purpose of the report; highlights the most important points from the analysis, such as the company's financial performance, competitive position, and growth prospects. \\
Company Overview & Offers a comprehensive description of the company, including its history, business model, and key products or services; discusses the company's organizational structure, management team, and corporate governance; analyzes the company's mission, vision, and strategic objectives. \\
Industry Analysis & Provides an in-depth analysis of the industry in which the company operates; includes information on market size, growth trends, and key drivers; discusses the regulatory environment, technological advancements, and other external factors affecting the industry; analyzes the industry's competitive dynamics and the company's position within the industry. \\
Competitive Landscape & Identifies the company's main competitors and their market share; compares the company's products, services, and pricing strategies with those of its competitors; analyzes the strengths and weaknesses of the company and its competitors; discusses potential new entrants and substitutes that could disrupt the competitive landscape. \\
Business Segments & Provides a detailed analysis of the company's various business segments or divisions; discusses the financial performance, growth prospects, and challenges of each segment; analyzes the contribution of each segment to the company's overall revenue and profitability. \\
Growth Strategies & Discusses the company's strategies for driving future growth, such as organic growth initiatives, product innovations, and geographic expansions; analyzes the company's mergers and acquisitions (M\&A) strategy and potential targets; examines the company's investments in research and development (R\&D) and marketing. \\
\midrule
\multicolumn{2}{l}{\textbf{Financial Analysis}} \\
Income Statement Analysis & Analyzes the company's revenue, expenses, and profitability. \\
Balance Sheet Analysis & Examines the company's assets, liabilities, and shareholders' equity. \\
Cash Flow Analysis & Analyzes the company's cash inflows and outflows to evaluate liquidity. \\
Financial Ratios & Discusses key ratios like profitability, liquidity, and solvency ratios. \\
\midrule
\multicolumn{2}{l}{\textbf{Strategic Outlook}} \\
Investment Thesis & Summarizes the key reasons for investing (or not investing) in the company's shares; discusses the potential catalysts and risks that could impact the company's valuation and stock price performance; provides a target price or price range for the company's shares based on the valuation analyses and investment thesis. \\
Valuation & Estimates the intrinsic value of the company's shares using various valuation methodologies, such as discounted cash flow (DCF) analysis, relative valuation multiples, and sum-of-the-parts analysis; compares the company's valuation with that of its peers and historical benchmarks; discusses the key assumptions and sensitivities underlying the valuation analyses. \\
\midrule
\multicolumn{2}{l}{\textbf{Risk and Governance}} \\
Risk Factors & Identifies and analyzes the key risks facing the company, such as market risks, operational risks, financial risks, and legal/regulatory risks; discusses the potential impact of these risks on the company's financial performance and growth prospects; examines the company's risk management strategies and mitigation measures. \\
Management and Governance & Provides an overview of the company's management team, including their experience, expertise, and track record; analyzes the company's corporate governance practices, such as board composition, executive compensation, and shareholder rights; discusses the company's succession planning and key person risks. \\
\bottomrule
ESG & Analyzes the company's performance and initiatives related to environmental sustainability, social responsibility, and corporate governance; discusses the potential impact of ESG factors on the company's reputation, risk profile, and financial performance; examines the company's compliance with ESG regulations and industry standards. \\
\midrule
\multicolumn{2}{l}{\textbf{Additional Content}} \\
Appendices and Disclosures & Includes additional supporting materials, such as financial statements, ratio calculations, and detailed segment data; provides important disclosures, such as the analyst's rating system, potential conflicts of interest, and disclaimers; discusses the limitations and uncertainties of the analysis and the need for further due diligence by investors. \\
None of the Above & Covers any content that does not fall into the specified topics. \\
\end{longtable}

\newpage
\begin{table}[!htb]
\caption{Turnover and Transaction Cost}
\label{tab:txcost_robust}
\footnotesize
{\footnotesize
This table reports monthly portfolio turnovers and net-transaction cost performances for long-short (L--S) strategies reported in Table \ref{tab:decile} using different lookback windows (LB). Turnover is measured as the average trading volume expressed as a percentage of portfolio value per month, reported separately for the long leg, short leg, and the combined L--S portfolio. Net returns account for proportional transaction costs and are computed as $R_{\text{net}} = R_{\text{gross}} - C \times \text{Turnover}_{L\text{--}S}$, where $C$ denotes the proportional transaction cost set to either 35 or 60 basis points. Appendix B details the turnover measurement and transaction cost implementation. The 35 basis point cost reflects trading expenses for large-cap stocks, while the 60 basis point cost corresponds to small-cap trading costs \citep{demiguel2020transaction}. t-statistics for net returns are reported in parentheses. ***, **, and * indicate statistical significance at the 1\%, 5\%, and 10\% levels, respectively.}
\begin{center}
\begin{tabularx}{\textwidth}{@{\hskip\tabcolsep\extracolsep\fill}lYYYYYYY}
\toprule
& \multicolumn{3}{c}{Turnover (\%)} & \multicolumn{1}{c}{Gross Mean} & \multicolumn{2}{c}{Net Mean (after costs)} \\
\cmidrule(lr){2-4}\cmidrule(lr){6-7}
LB & \multicolumn{1}{c}{Long} & \multicolumn{1}{c}{Short} & \multicolumn{1}{c}{L--S} & \multicolumn{1}{c}{(per mo)} & \multicolumn{1}{c}{$C=35$ bps} & \multicolumn{1}{c}{$C=60$ bps} \\
\midrule
9  & 26.4 & 41.6 & 34.0 & 0.98  & 0.751\sym{**}  & 0.579\sym{*}  \\
   &      &      &      &       & (2.20)          & (1.70)        \\
12 & 20.7 & 35.4 & 28.1 & 1.04  & 0.853\sym{**}  & 0.712\sym{**} \\
   &      &      &      &       & (2.51)          & (2.09)        \\
18 & 17.5 & 30.0 & 23.7 & 0.87  & 0.704\sym{**}  & 0.584\sym{*}  \\
   &      &      &      &       & (2.23)          & (1.85)        \\
24 & 13.7 & 23.2 & 18.4 & 1.16  & 1.040\sym{***} & 0.947\sym{***} \\
   &      &      &      &       & (3.10)          & (2.82)        \\
\bottomrule
\end{tabularx}
\end{center}
\end{table}

\newpage
\begin{landscape}
\begin{table}[!htb]
\caption{Correlations with Traded Factors and Industry Portfolios}
\scriptsize
\label{tab:corr}
    {\footnotesize This table reports the correlations between the report-based portfolio and traded factors in Panel A and 12 industry portfolios in Panel B. ``RP'' is the analyst narrative-based portfolio return; ``MKT-RF, SMB, HML, RMW, CMA, MOM'' are \citet{fama2018choosing} six factors; ``\text{R\_ME}, \text{R\_IA}, \text{R\_ROE}, \text{R\_EG}'' are \citet{hou2015digesting} factors; ``PEAD'' and ``FIN'' are \citet{daniel2020short} behavioral factors. The sample for traded factors and industry portfolios is from January 2005 to December 2023. Correlations are in percent.}
\begin{center}
\begin{tabularx}{1.4\textwidth}{lYYYYYYYYYYYYY}
\toprule
\multicolumn{14}{l}{Panel A: Correlations with Traded Factors} \\
\midrule
 & RP & \mbox{MKT-RF} & SMB & HML & RMW & CMA & MOM & \text{R\_ME} & \text{R\_IA} & \text{R\_ROE} & \text{R\_EG} & PEAD & FIN \\
\midrule
RP &   & 24.7 & -3.3 & -53.6 & 0.5 & -58.4 & -5.9 & -8.3 & -59.1 & -8.8 & 21.3 & 17.9 & -46.4 \\
MKT-RF & 24.7 &   & 38.0 & 15.3 & -18.6 & -12.4 & -38.9 & 40.8 & -12.0 & -42.4 & -40.2 & -30.0 & -31.7 \\
SMB & -3.3 & 38.0 &   & 30.4 & -38.4 & 4.2 & -29.8 & 95.8 & 7.3 & -53.5 & -57.5 & -25.2 & -34.5 \\
HML & -53.6 & 15.3 & 30.4 &   & 0.5 & 59.3 & -32.9 & 36.2 & 62.7 & -18.7 & -50.2 & -34.3 & 40.6 \\
RMW & 0.5 & -18.6 & -38.4 & 0.5 &   & 10.2 & 10.2 & -35.2 & 7.8 & 57.0 & 39.8 & 7.6 & 64.6 \\
CMA & -58.4 & -12.4 & 4.2 & 59.3 & 10.2 &   & 0.1 & 8.6 & 93.1 & 4.5 & -29.0 & -18.5 & 50.6 \\
MOM & -5.9 & -38.9 & -29.8 & -32.9 & 10.2 & 0.1 &   & -26.8 & -4.5 & 58.7 & 38.3 & 52.7 & 4.7 \\
\text{R\_ME} & -8.3 & 40.8 & 95.8 & 36.2 & -35.2 & 8.6 & -26.8 &   & 12.2 & -41.2 & -53.6 & -29.1 & -27.8 \\
\text{R\_IA} & -59.1 & -12.0 & 7.3 & 62.7 & 7.8 & 93.1 & -4.5 & 12.2 &   & -0.3 & -30.4 & -22.7 & 52.8 \\
\text{R\_ROE} & -8.8 & -42.4 & -53.5 & -18.7 & 57.0 & 4.5 & 58.7 & -41.2 & -0.3 &   & 62.0 & 33.9 & 46.2 \\
\text{R\_EG} & 21.3 & -40.2 & -57.5 & -50.2 & 39.8 & -29.0 & 38.3 & -53.6 & -30.4 & 62.0 &   & 38.1 & 23.5 \\
PEAD & 17.9 & -30.0 & -25.2 & -34.3 & 7.6 & -18.5 & 52.7 & -29.1 & -22.7 & 33.9 & 38.1 &   & -2.3 \\
FIN & -46.4 & -31.7 & -34.5 & 40.6 & 64.6 & 50.6 & 4.7 & -27.8 & 52.8 & 46.2 & 23.5 & -2.3 &   \\
\midrule
\end{tabularx}
\begin{tabularx}{1.4\textwidth}{lYYYYYYYYYYYYY}
\multicolumn{13}{l}{Panel B: Correlations with Industry Portfolios} \\
\midrule
 & RP & Nodur & Durbl & Manuf & Enrgy & Chems & Buseq & Telcm & Utils & Shops & Hlth & Money & Other \\
\midrule
RP &   & 0.5 & 33.5 & 17.1 & -3.3 & 12.7 & 45.1 & 8.8 & -1.0 & 28.7 & 9.4 & -0.2 & 15.6 \\
Nodur & 0.5 &   & 52.3 & 75.4 & 54.2 & 84.2 & 66.9 & 76.4 & 65.6 & 70.5 & 69.1 & 70.3 & 77.8 \\
Durbl & 33.5 & 52.3 &   & 73.7 & 44.1 & 62.4 & 75.7 & 57.2 & 33.4 & 74.6 & 49.8 & 68.2 & 77.0 \\
Manuf & 17.1 & 75.4 & 73.7 &   & 66.7 & 86.8 & 82.3 & 78.6 & 54.7 & 79.2 & 68.7 & 84.7 & 93.8 \\
Enrgy & -3.3 & 54.2 & 44.1 & 66.7 &   & 57.2 & 48.2 & 57.4 & 46.3 & 44.9 & 43.7 & 58.5 & 61.3 \\
Chems & 12.7 & 84.2 & 62.4 & 86.8 & 57.2 &   & 75.3 & 75.2 & 60.4 & 75.3 & 70.5 & 75.8 & 86.3 \\
Buseq & 45.1 & 66.9 & 75.7 & 82.3 & 48.2 & 75.3 &   & 69.8 & 48.4 & 83.9 & 67.0 & 71.6 & 84.2 \\
Telcm & 8.8 & 76.4 & 57.2 & 78.6 & 57.4 & 75.2 & 69.8 &   & 56.6 & 71.5 & 64.5 & 73.4 & 79.7 \\
Utils & -1.0 & 65.6 & 33.4 & 54.7 & 46.3 & 60.4 & 48.4 & 56.6 &   & 47.1 & 50.1 & 41.5 & 53.2 \\
Shops & 28.7 & 70.5 & 74.6 & 79.2 & 44.9 & 75.3 & 83.9 & 71.5 & 47.1 &   & 70.2 & 74.6 & 85.2 \\
Hlth & 9.4 & 69.1 & 49.8 & 68.7 & 43.7 & 70.5 & 67.0 & 64.5 & 50.1 & 70.2 &   & 67.1 & 71.8 \\
Money & -0.2 & 70.3 & 68.2 & 84.7 & 58.5 & 75.8 & 71.6 & 73.4 & 41.5 & 74.6 & 67.1 &   & 88.3 \\
Other & 15.6 & 77.8 & 77.0 & 93.8 & 61.3 & 86.3 & 84.2 & 79.7 & 53.2 & 85.2 & 71.8 & 88.3 &   \\
\bottomrule
\end{tabularx}
\end{center}
\end{table}
\end{landscape}

\newpage
\begin{table}[!htb]
\footnotesize
\caption{Sentiment Portfolio Statistics}
\label{tab:sentport}
    {\footnotesize This table reports the value-weighted decile portfolio performances sorted by the average of four measures of the past 12 months' analyst reports. Contemporaneous Return denotes two-day DGTW characteristic-adjusted buy-and-hold abnormal returns following report release dates. Recommendation Revision denotes recommendation revision, calculated as the current report's recommendation minus the last recommendation in I/B/E/S issued by the same analyst for the same stock. Headline Sentiment represents the sentiment score of report headlines measured using a fine-tuned BERT model. Body Sentiment represents the average sentiment score of report body content measured using a fine-tuned BERT model. For each portfolio, I report several performance measures: excess return mean and standard deviation, Sharpe ratio, $\alpha$ relative to \citet{fama2018choosing} six factors, and the corresponding t-statistics for $\alpha$. Portfolios are rebalanced monthly. The `H-L' row represents a long-short strategy that takes a long position in the highest decile (High) and a short position in the lowest decile (Low). The sample period extends from January 2005 to June 2024.}
    \begin{center}
    \scalebox{1}{
    \begin{tabularx}{\textwidth}{@{\hskip\tabcolsep\extracolsep\fill}l*{10}r}
    \toprule
    & \multicolumn{5}{c}{Contemporaneous Return} & \multicolumn{5}{c}{Recommendation Revision} \\
    \cmidrule{2-6} \cmidrule{7-11}
     & Mean & SD & SR & $\alpha$ & $t_\alpha$  & Mean & SD & SR & $\alpha$ & $t_\alpha$ \\
    \midrule
    Low (L) & 1.10 & 6.66 & 0.57 & 0.26 & 1.42 & 0.84 & 4.70 & 0.62 & 0.01 & 0.11 \\
    2 & 0.93 & 5.99 & 0.54 & 0.14 & 0.86 & 0.87 & 4.81 & 0.62 & 0.09 & 0.95 \\
    3 & 0.71 & 5.19 & 0.47 & -0.01 & -0.10 & 0.88 & 4.61 & 0.66 & 0.10 & 0.75 \\
    4 & 0.82 & 4.45 & 0.64 & 0.11 & 1.21 & 0.46 & 4.30 & 0.37 & -0.33 & -3.63 \\
    5 & 0.80 & 4.53 & 0.61 & -0.08 & -0.93 & 0.54 & 4.90 & 0.38 & -0.20 & -1.73 \\
    6 & 0.79 & 4.41 & 0.62 & 0.00 & 0.03 & 0.69 & 5.31 & 0.45 & -0.12 & -1.07 \\
    7 & 0.84 & 4.56 & 0.64 & 0.00 & 0.05 & 0.81 & 5.61 & 0.50 & -0.04 & -0.29 \\
    8 & 0.94 & 4.78 & 0.68 & 0.08 & 0.75 & 1.02 & 4.90 & 0.72 & 0.15 & 1.32 \\
    9 & 1.12 & 5.43 & 0.72 & 0.27 & 2.02 & 0.77 & 4.54 & 0.59 & 0.01 & 0.08 \\
    High (H) & 0.78 & 6.06 & 0.44 & -0.10 & -0.72 & 1.11 & 4.73 & 0.81 & 0.25 & 1.94 \\
    H-L & -0.33 & 4.20 & -0.27 & -0.36 & -1.57 & 0.27 & 2.68 & 0.35 & 0.24 & 1.56 \\
    \end{tabularx}}
    \vspace{0.3in}   
    \scalebox{1
    }{
    \begin{tabularx}{\textwidth}{@{\hskip\tabcolsep\extracolsep\fill}l*{10}r}
    \toprule
    & \multicolumn{5}{c}{Headline Sentiment} & \multicolumn{5}{c}{Body Sentiment} \\
    \cmidrule{2-6} \cmidrule{7-11}
     & Mean & SD & SR & $\alpha$ & $t_\alpha$  & Mean & SD & SR & $\alpha$ & $t_\alpha$ \\
    \midrule
    Low (L) & 0.48 & 5.57 & 0.30 & -0.23 & -2.02 & 0.98 & 5.75 & 0.59 & 0.22 & 1.25 \\
    2 & 0.63 & 5.60 & 0.39 & -0.09 & -0.97 & 0.48 & 5.47 & 0.30 & -0.17 & -1.83 \\
    3 & 0.81 & 5.16 & 0.55 & 0.07 & 0.60 & 0.84 & 5.18 & 0.56 & 0.14 & 1.41 \\
    4 & 0.98 & 4.58 & 0.74 & 0.25 & 2.65 & 0.65 & 4.80 & 0.47 & -0.09 & -0.88 \\
    5 & 1.02 & 4.57 & 0.78 & 0.18 & 1.84 & 1.02 & 4.96 & 0.71 & 0.18 & 1.63 \\
    6 & 0.80 & 4.55 & 0.61 & -0.02 & -0.17 & 0.87 & 4.71 & 0.64 & 0.05 & 0.41 \\
    7 & 0.78 & 4.77 & 0.57 & -0.08 & -0.97 & 0.81 & 4.75 & 0.59 & -0.10 & -1.35 \\
    8 & 0.67 & 4.55 & 0.51 & -0.19 & -1.68 & 0.82 & 4.59 & 0.62 & -0.03 & -0.31 \\
    9 & 0.85 & 4.93 & 0.60 & -0.03 & -0.19 & 0.90 & 4.50 & 0.70 & 0.03 & 0.29 \\
    High (H) & 0.74 & 5.11 & 0.50 & -0.15 & -1.50 & 0.82 & 4.56 & 0.63 & -0.01 & -0.16 \\
    H-L & 0.27 & 3.31 & 0.28 & 0.08 & 0.58 & -0.16 & 4.17 & -0.13 & -0.24 & -0.97 \\
    \bottomrule
    \end{tabularx}}
    \end{center}
\end{table}

\newpage
\begin{table}[!htb]
\caption{Incremental Investment Value: Strategic Outlook}
\footnotesize
\label{tab:IIV_SO}
    {\footnotesize This table examines the incremental investment value of analyst reports compared to existing factors. The analysis includes: strategic outlook content (SO), 18 analyst-based factors from \cite{ChenZimmermann2021} (ANA), and 92 fundamental-based factors from \cite{gu2020empirical} (ANOM). For each strategy and combination, I report mean returns, Sharpe ratio (SR), and information ratio (IR), with corresponding t-statistics in parentheses. I also report the alphas of long-short portfolios relative to four benchmark factor models: \citet{fama2015five} five factors ($\alpha_{F5}$), \citet{fama2018choosing} six factors ($\alpha_{F6}$), \citet{hou2015digesting} factors ($\alpha_{HXZ}$), and \citet{daniel2020short} factors ($\alpha_{DHS}$). The sample for traded factors is from January 2005 to December 2023.}
    \begin{center}
\begin{tabularx}{\textwidth}{lYYYYYYY}
\toprule
 & Mean & SR & $\alpha_{F5}$ & $\alpha_{F6}$ & $\alpha_{HXZ}$ & $\alpha_{DHS}$ & IR \\
\midrule
\multicolumn{8}{l}{Panel A: Factor Performances} \\
\midrule
SO & 1.30 & 0.87 & 0.95 & 0.96 & 1.04 & 1.46 & - \\
 & (3.79) & (3.74) & (3.03) & (3.01) & (3.50) & (3.89) & - \\
ANA & 0.27 & 0.48 & 0.28 & 0.25 & 0.22 & 0.16 & - \\
 & (2.08) & (2.07) & (2.80) & (3.01) & (2.08) & (1.26) & - \\
ANOM & 1.34 & 1.03 & 1.40 & 1.31 & 1.28 & 1.15 & - \\
 & (4.47) & (4.38) & (5.05) & (4.65) & (4.43) & (3.84) & - \\
ANA + ANOM & 0.80 & 1.02 & 0.84 & 0.78 & 0.75 & 0.66 & - \\
 & (4.44) & (4.35) & (5.30) & (5.16) & (4.36) & (3.58) & - \\
\midrule
\multicolumn{8}{l}{Panel B: Strategic Outlook versus Factors} \\
\midrule
SO + ANA & 0.78 & 1.07 & 0.62 & 0.61 & 0.63 & 0.81 & 0.96 \\
 & (4.68) & (4.57) & (4.34) & (4.26) & (5.30) & (4.95) & (3.45) \\
SO + ANOM & 1.32 & 1.69 & 1.17 & 1.14 & 1.16 & 1.30 & 1.34 \\
 & (7.37) & (6.97) & (7.92) & (7.79) & (8.21) & (7.92) & (4.24) \\
SO + ANA + ANOM & 0.97 & 1.74 & 0.88 & 0.84 & 0.84 & 0.92 & 1.41 \\
 & (7.57) & (7.14) & (9.05) & (9.55) & (9.51) & (8.64) & (4.35) \\
\bottomrule
\end{tabularx}
\end{center}
\end{table}

\begin{table}[!htb]
\caption{Post-Knowledge Cutoff Portfolio Performance: Strategic Outlook}
\footnotesize
\label{tab:lab_so}
{\footnotesize This table reports value-weighted decile portfolio performance during post-language-model training knowledge-cutoff periods. Portfolios are formed using each model’s embeddings of \emph{strategic outlook} content. Reported are the monthly mean return (Mean), standard deviation (SD), and Sharpe ratio (SR). The analysis spans the following knowledge-cutoff-based windows: $\text{ChronoGPT}_{1999}$ and $\text{ChronoGPT}_{2024}$ are evaluated across the entire out-of-sample, January 2005–June 2024, to enable an apple-for-apple comparison of the same architecture with different knowledge cutoffs. BERT (January 2019–June 2024), RoBERTa (March 2019–June 2024), LLaMA2-13B (October 2022–June 2024), and LLaMA3-8B (April 2023–June 2024). ``H--L'' is a long–short strategy that buys the highest decile (High) and shorts the lowest decile (Low).}
\begin{center}
\begin{tabularx}{\textwidth}{lYYYYYYYYY}
\toprule
& \multicolumn{3}{c}{$\text{ChronoGPT}_{1999}$} & \multicolumn{3}{c}{$\text{ChronoGPT}_{2024}$} & \multicolumn{3}{c}{BERT} \\
\cmidrule(lr){2-4} \cmidrule(lr){5-7} \cmidrule(lr){8-10}
 & Mean & SD & SR & Mean & SD & SR & Mean & SD & SR \\
\midrule
Low (L) & 0.63 & 4.18 & 0.52 & 0.68 & 4.27 & 0.55 & 0.82 & 5.69 & 0.50 \\
2       & 0.71 & 4.45 & 0.55 & 0.73 & 4.46 & 0.57 & 1.10 & 4.92 & 0.78 \\
3       & 0.66 & 4.62 & 0.49 & 0.58 & 4.68 & 0.43 & 0.92 & 5.16 & 0.61 \\
4       & 0.66 & 4.49 & 0.46 & 0.73 & 4.76 & 0.53 & 1.39 & 5.04 & 0.95 \\
5       & 0.78 & 4.78 & 0.57 & 0.81 & 5.45 & 0.56 & 1.20 & 5.15 & 0.81 \\
6       & 0.93 & 4.68 & 0.69 & 0.95 & 4.86 & 0.67 & 0.66 & 5.52 & 0.42 \\
7       & 0.84 & 5.11 & 0.57 & 0.94 & 4.89 & 0.66 & 1.14 & 5.48 & 0.72 \\
8       & 0.88 & 4.90 & 0.62 & 0.91 & 5.11 & 0.61 & 1.53 & 5.77 & 0.92 \\
9       & 0.79 & 5.74 & 0.48 & 0.81 & 5.36 & 0.52 & 2.17 & 6.67 & 1.13 \\
High (H)& 1.55 & 6.35 & 0.84 & 1.56 & 6.62 & 0.84 & 3.19 & 8.09 & 1.37 \\
H--L    & 0.91 & 4.87 & 0.65 & 0.88 & 4.75 & 0.64 & 2.37 & 6.51 & 1.26 \\
\midrule
& \multicolumn{3}{c}{RoBERTa} & \multicolumn{3}{c}{LLaMA2} & \multicolumn{3}{c}{LLaMA3} \\
\cmidrule(lr){2-4} \cmidrule(lr){5-7} \cmidrule(lr){8-10}
 & Mean & SD & SR & Mean & SD & SR & Mean & SD & SR \\
\midrule
Low (L) & 0.97 & 5.36 & 0.63 & 1.08 & 4.23 & 0.89 & 0.01 & 3.48 & 0.01 \\
2       & 0.77 & 5.04 & 0.53 & 1.21 & 4.37 & 0.96 & 1.19 & 3.77 & 1.09 \\
3       & 0.98 & 5.22 & 0.65 & 1.15 & 4.51 & 0.88 & 1.05 & 4.19 & 0.87 \\
4       & 0.99 & 4.96 & 0.69 & 1.39 & 4.49 & 1.07 & 2.05 & 4.97 & 1.43 \\
5       & 0.68 & 5.37 & 0.44 & 1.82 & 5.01 & 1.26 & 1.69 & 3.60 & 1.63 \\
6       & 0.95 & 5.34 & 0.62 & 1.91 & 4.85 & 1.36 & 0.66 & 4.56 & 0.51 \\
7       & 1.35 & 5.04 & 0.93 & 1.45 & 4.73 & 1.06 & 0.67 & 5.09 & 0.46 \\
8       & 1.34 & 5.56 & 0.84 & 2.35 & 5.64 & 1.44 & 1.77 & 4.81 & 1.28 \\
9       & 1.94 & 7.33 & 0.92 & 2.30 & 6.14 & 1.30 & 2.26 & 5.13 & 1.53 \\
High (H)& 2.78 & 8.28 & 1.16 & 3.73 & 7.10 & 1.82 & 5.29 & 7.88 & 2.32 \\
H--L    & 1.81 & 7.37 & 0.85 & 2.64 & 6.95 & 1.32 & 5.28 & 7.73 & 2.37 \\
\bottomrule
\end{tabularx}
\end{center}
\end{table}

\clearpage
\appendix

\section*{B. Turnover and Transaction Cost}\label{sec:turnover}
\small

\paragraph{Turnover Measurement.}
I follow \citet{demiguel2009optimal} in turnover measurement, which quantifies the trading intensity required for portfolio rebalancing. Specifically, monthly turnover ${\widehat{\boldsymbol{w}}}_{t}$ is measured as the $\ell_1$ norm of the adjustment needed to transition from passively evolved holdings ${\boldsymbol{w}^{\text{drift}}_{t+1}}$ to newly optimized target weights ${\widehat{\boldsymbol{w}}}_{t+1}$.

Let $\widetilde{\boldsymbol{R}}_{t+1}$ represent the vector of excess returns for individual stocks during period $t\!+\!1$, and let $r_{f,t+1}$ denote the risk-free rate over the same interval. In the absence of active rebalancing, the portfolio weights evolve passively according to:
\[
\boldsymbol{w}^{\text{drift}}_{t+1} = \frac{\widehat{\boldsymbol{w}}_{t}\,\circ\bigl(\mathbf{1}+r_{f,t+1}\mathbf{1}+\widetilde{\boldsymbol{R}}_{t+1}\bigr)}
{\,1+r_{f,t+1}+\widehat{\boldsymbol{w}}_{t}'\,\widetilde{\boldsymbol{R}}_{t+1}\,},
\]
where $\circ$ denotes the Hadamard (element-wise) product and the denominator normalizes by the total portfolio return to maintain unit investment.

The turnover for period $t\!+\!1$ is then defined as:
\[
\text{turnover}_{t+1}
\;=\;
\Biggl\|
\widehat{\boldsymbol{w}}_{t+1}
-
\boldsymbol{w}^{\text{drift}}_{t+1}
\Biggr\|_{1}
\;=\;
\sum_{i=1}^{N_{t+1}} \left|\widehat{w}_{i,t+1} - w^{\text{drift}}_{i,t+1}\right|.
\]

The empirical analysis in Table \ref{tab:txcost_robust} reports out-of-sample averages of monthly turnover rates.

\paragraph{Transaction Cost Implementation.}
To evaluate the economic significance of trading costs, I incorporate proportional transaction expenses into net return calculations. Given the model-implied gross excess return $R^{\text{gross}}_{t+1}$ and a proportional trading cost of $C$ per dollar traded, the net excess return after transaction costs is computed as:
\[
R^{\text{net}}_{t+1}
\;=\;
\bigl(1+r_{f,t+1}+R^{\text{gross}}_{t+1}\bigr)
\bigl(1-C\cdot\text{turnover}_{t+1}\bigr)
-
\bigl(1+r_{f,t+1}\bigr).
\]

This specification assumes that transaction costs are proportional to the dollar volume of trades and are incurred symmetrically for both buying and selling. I consider cost parameters with $C=0.006$ (60 basis points) and $C=0.0035$ (35 basis points), following the specifications in \citet{demiguel2020transaction}.

\end{document}